\documentclass[a4paper,11pt]{article}
\usepackage{graphicx}
\usepackage{xcolor}
\usepackage{amsmath,amsfonts,amssymb,amstext,hyperref,hypcap}
\usepackage{multirow}
\usepackage{cancel}
\usepackage{empheq}
\usepackage{float}
\usepackage{cite}
\usepackage{subcaption}
\usepackage{authblk}
\usepackage{comment}
\usepackage{blindtext}
\usepackage{cases}
\usepackage{ulem}
\usepackage{nccmath, mathtools}
\usepackage{mleftright}
\usepackage{epsfig}%
\usepackage[bottom]{footmisc}

\usepackage{epsfig}%
\usepackage[bottom]{footmisc}
\begin{document}
\date{}

\title{\centerline \bf  
Exploring the evolution of structure growth in the universe with field-fluid interactions through dynamical stability analysis}
\bigskip
\author[1]{ Anirban Chatterjee\thanks{Corresponding author: iitkanirbanc@gmail.com, anirbanchatterjee@nbu.edu.cn}}
 
\author[2]{Abhijit Bandyopadhyay\thanks{abhijit.phy@gm.rkmvu.ac.in}}
\author[2]{Debasish Majumdar\thanks{debasish.majumdar@gm.rkmvu.ac.in}}
\affil[1]{Institute of Fundamental Physics and Quantum Technology,\par
 Department of Physics, School of Physical Science and Technology,\par 
  Ningbo University, Ningbo, Zhejiang 315211, China}\par

\affil[2]{Department of Physics, Ramakrishna Mission Vivekananda Educational and Research Institute, Belur Math, Howrah-711202, West-Bengal, India}
 
\date{\today}
\maketitle
 
\begin{abstract} 
We investigate an interacting quintessence dark energy - dark matter scenario 
and its impact on
structure formation by analyzing the evolution of scalar perturbations.  
The interaction is 
introduced by incorporating a non-zero source term into the continuity equations of the two 
sectors (with opposite signs), modeled as $\bar{Q}_0 \equiv \alpha\bar{\rho}_{\rm m}(H + 
\kappa\dot{\phi})$.
The coupling parameter $\alpha$ and the parameter $\lambda$ involved in quintessence potential  
$V(\phi) = V_0e^{-\lambda\kappa\phi}$, play crucial roles in governing the dynamics of 
evolution examined within the present framework. The cosmic evolution, within this context, is 
depicted as a first-order autonomous system of equations involving appropriately chosen 
dynamical variables.
We analyzed the associated stability characteristics and growth rate of perturbations,  and 
obtained domains in the ($\alpha-\lambda$) parameter space for which fixed points can exhibit 
stable and non-phantom accelerating solutions. 
Depending on its magnitude, the coupling parameter  $\alpha$ has the potential to change the characteristics of certain critical points, altering them from attractors to repellers.
 This model effectively captures the evolutionary features of the universe across its 
various phases at both the background and perturbation levels. The issue of cosmic coincidence 
can also be addressed within the framework of this model. We also observed 
that for a moderate strength of coupling, the growth rate of matter 
perturbation  
extends into the distant future. 
\end{abstract}

\section{Introduction}
 Cosmological observations over the last few decades have revealed that the universe is currently undergoing accelerated expansion, and the transition from a decelerating phase to the current accelerated phase occurred during the late stage of cosmic evolution. The initial empirical evidence instrumental in establishing the fact came from the interpretation of luminosity distance and redshift measurements of type Ia supernovae (SNe Ia) events \cite{ref:Riess98, ref:Perlmutter}.
Further support for this late-time cosmic acceleration has been provided through the examination of temperature fluctuations in the cosmic microwave background using WMAP \cite{WMAP:2003elm,Hinshaw:2008kr}, the observation of baryon acoustic oscillations \cite{SDSS:2005xqv}, and the analysis of the power spectrum of matter distributions in the universe \cite{mps1,mps2}. This late-time cosmic acceleration is attributed to `Dark Energy' (DE), a hypothetical component exerting negative pressure to counteract gravitational attraction and propel the acceleration. Nevertheless, comprehension of nature and origin of DE remains a prominent unresolved enigma in contemporary cosmology.  
In addition to dark energy, which constitutes around 70\% of the universe, the matter content of the universe, besides accounting for the known luminous or baryonic matter, consists of a large amount of unknown nonluminous matter whose presence is known only through its gravitational influence. This includes flattened rotation curves of galaxies \cite{Sofue:2000jx}, gravitational lensing \cite{Bartelmann:1999yn} and microlensing, 
the Bullet cluster, and other colliding clusters phenomena \cite{Clowe:2006eq}.
 The amount of this unknown nonluminous matter, known as `Dark Matter' (DM), 
 is estimated to be around 26\% of the total content of the universe, with the known baryonic matter amounting to only about a meager 
4\%. This energy budget of the universe has been estimated from observational measurements of the PLANCK satellite-based experiment \cite{Planck:2018vyg}. \\

Various theoretical perspectives have emerged, 
each aiming to formulate DE models that elucidate the observed cosmic acceleration.   One of the popular phenomenological models is the $\Lambda$-CDM model, where Cold Dark Matter (CDM) is regarded as the dominant matter component of the universe, and cosmic acceleration is generated by 
exploiting the cosmological constant $\Lambda$ introduced in the Einstein field equation as $\Lambda g_{\mu\nu}$. 
Though the value of $\Lambda$  can be tuned 
  to align the model predictions  with the observed features of cosmic acceleration, it grapples with issues like the coincidence \cite{Zlatev:1998tr} and fine-tuning problems \cite{Martin:2012bt}. This motivates exploration of alternative models for DE. One subset of these DE models encompasses field-theoretic models, which involve modification of the energy-momentum tensor in Einstein's field equations due to the presence of a (scalar) field as a  component of the universe, distinct from matter and radiation. Examples of such models include both quintessence   \cite{Peccei:1987mm,Ford:1987de,Peebles:2002gy,Nishioka:1992sg, Ferreira:1997au,Ferreira:1997hj,Caldwell:1997ii,Carroll:1998zi,Copeland:1997et} and $k$-essence models \cite{Fang:2014qga, ArmendarizPicon:1999rj,ArmendarizPicon:2000ah,ArmendarizPicon:2000dh,ArmendarizPicon:2005nz,Chiba:1999ka,ArkaniHamed:2003uy,Caldwell:1999ew,Bandyopadhyay:2017igc,Bandyopadhyay:2018zlz,Bandyopadhyay:2019ukl,Bandyopadhyay:2019vdd,Chatterjee:2022uyw}. 
  The scalar field serves the purpose of generating the required negative pressure to propel cosmic acceleration, achieved either through the slowly varying potentials in quintessence models or by harnessing its kinetic energy in $k$-essence models. 
   Another class of DE models involves modification of the geometric part of Einstein's equations, particularly the Einstein-Hilbert action, to generate
    late-time cosmic acceleration. These models include $f(R)$ gravity models, scalar-tensor theories, Gauss-Bonnet gravity, and braneworld models of dark energy, as documented in references \cite{fr1,fr2,fr3,fr4,fr5,fr6,fr7,fr8,fr9}.\\

A multitude of comprehensive studies already exist, examining the potential dynamics 
and evolution of dark energy within different  DE   models. These investigations 
assume the independent evolution of the DE component solely through its coupling to 
gravity, without interactions with matter field(s). Recently, there has been an 
increasing interest, from various perspectives, in exploring scenarios that 
incorporate interactions between DE and  DM  fields.
In principle, there is no reason to exclude considerations of  DE-DM  interactions. 
The fundamental requirement of conserving the total energy-momentum tensor
($T^{\mu\nu}$)  for the universe ($\nabla_\mu T^{\mu\nu} = 0$) can be 
trivially satisfied by conserving the energy-momentum tensor of individual components. 
This requirement can also be fulfilled by setting $\nabla_\mu T_i^{\mu\nu} = Q_i^\nu 
\neq 0$ for each individual component (labeled as $i$), subject to $\sum_i Q_i^\nu = 
0$. The presence of non-vanishing $Q_i^\nu$ implies the existence of interactions 
between different components.
Remarkably, the interaction between the DE and DM sectors naturally emerges when 
establishing the equivalence between modified gravity theories and scalar-tensor 
theories of dark energy through conformal transformations 
\cite{mgst1,mgst2,mgst3,mgst4}.  
Moreover, selecting an appropriate interaction between a scalar field and a matter fluid proves to be successful in addressing the coincidence problem \cite{Chimento:2003iea}  of cosmology.  
Investigating the impacts of interactions involving coupled quintessence dark energy on the Cosmic Microwave Background (CMB) and matter power spectrum has the potential to reconcile the tension between observations of the CMB and the inferred structure growth from cluster counts. 
Interacting DE-DM models can provide
 potential resolutions to the problem of  
  $4$-$5\sigma$ level discrepancy between the measured value of 
Hubble parameter at present epoch  ($H_0$ )
by Planck collaboration    
and its locally measured value by the SH0ES collaboration   
(the Hubble tension problem)
\cite{ht1,ht2,Pourtsidou:2016ico,An:2017crg,Kumar:2017dnp}.
Also, the DE-DM interaction scenarios are more favored than the $\Lambda$-CDM model, as revealed by the combined analysis of cosmic shear data from the Kilo Degree Survey (KiDS) \cite{kids} and angular power spectra from Planck's cosmic microwave background measurements \cite{plxi}. The significant discordance between these two datasets, as interpreted by the $\Lambda$-CDM model, is mitigated in scenarios involving interacting DE-DM \cite{An:2017crg}.
The discrepancy between the high $\sigma_8$ value (which quantifies the growth of matter fluctuations in the late universe) estimated by Planck assuming $\Lambda$-CDM and the lower value preferred by cosmic shear measurements is also alleviated in interacting scenarios \cite{Yang:2018uae,DiValentino:2021izs}. \\

The primary objective of this study is to analyze the impact of DE-DM interactions on structure formation in the universe by examining the gravitational evolution of scalar perturbations over a flat spacetime background.
Although there  exist 
quantum field theory-inspired models that give rise to interactions between 
DE and DM sectors \cite{qft1,qft2},
we employ a phenomenological model to describe the interactions between DE and DM. 
The  investigation of the evolution of the universe, incorporating DE-DM interactions, 
first begins with an analysis at the background level. At this stage, the universe is characterized by the flat Friedmann-Lema\^itre-Robertson-Walker (FLRW) metric.
We depict the interaction by introducing a non-zero source 
terms on the right side of the  continuity
 equations for both  DE and  DM  sectors. 
These terms are parametrized in terms of energy densities 
of both dark fluids at the background level, 
all while ensuring overall energy-momentum conservation. 
However, distinguishing among the various alternative models of 
DE-DM interactions require an investigation beyond background evolution introducing inhomogeneities
and following their evolution. This entails consideration of 
theory of linear cosmological perturbations and 
investigating impact of   interactions on the growth of perturbations in the universe.  
Under the assumption of no anisotropic stress,
the scalar perturbations to the metric 
  in Newtonian gauge
can be completely characterised in terms of a single scalar 
function.   
As  growth of cosmic structure takes place on spatial scales considerably smaller than the horizon, our focus is solely on perturbations within the matter sector, disregarding perturbations in the  DE  sector. 
The progression of perturbations can then be described in terms of the evolution of  DM  density contrast ($\delta$) 
obtained using the continuity equation and the evolution
of divergence of velocity perturbations ($\theta$) in Fourier space, resulting from the Euler equation due to momentum conservation. 
Examining the DM perturbations entails combining the above equations   
utilizing the time-time Einstein field equation (Poisson equation).\\

To explore the impact of DE-DM interactions on the formation of structures 
through matter clustering, we treat DM as non-relativistic dust.  
We characterize DE  using a  (quintessence) scalar field  $\phi$ whose dynamics is driven by the Lagrangian   
${\cal L}_{\phi} = \frac12 \partial_\mu \phi\partial^\mu \phi +  V(\phi)$ with 
$V(\phi) = V_0 e^{-\lambda\kappa\phi}$,
where $V_0>0$ is a   constant and $\lambda$ represents a dimensionless parameter.
In principle, one may  incorporate interactions between DE and DM  sectors 
at a phenomenological level by modeling the non-zero source term ($\bar{Q}_0$) 
in the unperturbed continuity equations for both sectors. 
This involves expressing $\bar{Q}_0$ in terms of various cosmological quantities 
such as the unperturbed energy densities ($\bar{\rho}_{\rm m}$, $\bar{\rho}_\phi$) of the dark fluid components, 
the time derivative of the quintessence field ($\dot{\phi}$), Hubble parameter ($H$), FLRW scale factor ($a$)
\textit{etc}. Consideration of different functional forms of $\bar{Q}_0$ based on these quantities, while ensuring that $\bar{Q}_0$ has the proper mass dimension consistent with the terms in the continuity equation, facilitates a model-dependent and comprehensive exploration of interacting scenarios. 
Examining the implications of specific couplings, such as $ \bar{Q}_0 \sim \bar{\rho}_{\rm m} \dot{\phi}$ as investigated in \cite{Wetterich:1994bg,Amendola:1999er}, which is capable of producing stable late-time cosmic acceleration, or couplings like $\bar{Q}_0 \sim (\bar{\rho}_{\rm m}+\bar{\rho}_\phi)\dot{\phi}$, which have been demonstrated not to be viable in describing all three cosmological eras \cite{Bernardi:2016xmb}, has
 offered valuable insights in the context of dark energy-dark matter interaction scenarios.
Also, a coupling of the form $\sim \bar{\rho}_\phi\dot{\phi}$ would not induce any impact
on structure formation, 
as dark energy is predicted not to cluster at sub-horizon scales \cite{Duniya:2013eta}.\\

In this work, we adopt an interacting model characterised by 
$   \bar{Q}_0 \equiv \alpha \bar{\rho}_{\rm m}(H + \kappa \dot{\phi})$,
where  $\alpha$ decides the strength of the coupling.
In quintessence   models, late-time cosmic acceleration is achieved by suppressing 
the kinetic part ($\dot{\phi}^2$) of the Lagrangian with respect to 
the potential $V(\phi)$ ($\dot{\phi}^2 \ll V(\phi)$), 
causing the equation of state parameter to drop below $-\frac13$.
However, during the growth of matter perturbations in eras preceding DE domination,
 the kinetic term involving $\dot{\phi}$ is not insignificant. 
 The adopted expression for $\bar{Q}_0$ in this context depicts a scenario in which the 
 rate of energy exchange between DE and DM sectors due to interaction 
 is proportional to the product of DM density and the sum of two distinct temporal rates, 
 \textit{viz.}, the rate of change of $\phi$ and the expansion rate of the universe ($H$). 
In this paper, we analyse this model of interacting quintessence in conjunction with linear cosmological perturbations, 
employing the theory of dynamical systems. This type of analysis provides a scope for comparing stability, perturbation growth, and other distinctive features among various interacting models \cite{Marcondes:2016reb,Dutta:2017wfd, Khyllep:2021wjd}.\\

With the incorporation of DE-DM interactions,
the equations addressing evolution  of the universe, 
at the level of background   and perturbations 
exhibit severe complexities and pose
significant challenges in obtaining their
analytical or even numerical solutions. 
Employing the theory of dynamical systems and 
associated phase space analysis in such situations 
emerges as a robust mathematical tool for 
extracting analytical insights into the global behaviour of evolution
 circumventing the need for direct analytical or numerical
 approaches \cite{Marcondes:2016reb,Dutta:2017wfd, Khyllep:2021wjd,Chatterjee:2021ijw,Chatterjee:2021hhj,Hussain:2022osn,Bhattacharya:2022wzu,Hussain:2023kwk,Chatterjee:2023uga, Cabral:2009hoy, Tsujikawa:2012hv}.  
The dynamical analysis approach formulates the evolution equations as a set
of autonomous equations, expressed in terms of   dimensionless dynamical parameters,
appropriately defined in terms of various quantities relevant to the context.
The autonomous system corresponds to 
evolution of linear perturbations in a universe with DE-DM 
interactions characterised by $\bar{Q}_0 \equiv \alpha \bar{\rho}_{\rm m}(H + \kappa \dot{\phi})$,
turns out to be a 3-dimensional autonomous 
system realised in terms of time derivatives 
of three suitably  chosen   dynamical
variables $x$,$y$ and $u$ (see Sec.\ \ref{sec:DSA} for details)
conceived as a function of time parameter $N = \ln a$,
with $u$ being an indicator of 
rate of growth of perturbations.
A  phase space analysis of the system investigating its critical points offers insights into the asymptotic behavior of the model.
The analysis scenario involves two parameters \textit{viz.}   the DE-DM coupling  $\alpha$, and the constant $\lambda$   in the potential $V(\phi) (= V_0e^{-\lambda \kappa \phi})$
of the quintessence model. The critical points, in general, and their stability traits depend on these parameters.  
Moreover, the unique critical phases associated with the critical points can be inferred from the values of the density parameters for various components and the equation of state for the overall fluid at these critical points. 
The dynamical analysis yields interesting cosmological insights at both the background and perturbation levels, highlighting the significant influence of DE-DM interaction on both background evolution and structure growth. 
Such analysis facilitates the identification of growing mode trajectories regardless of particular initial conditions, and also examines the evolution of matter perturbations across distinct cosmological epochs defined by each critical point. The dynamical analysis reveals  interesting cosmology 
at both background and perturbation levels indicating
significant impact of
 DE-DM interaction  on both background evolution and 
structure growth. This approach enables the determination of the
growing mode trajectories independently of specific initial
conditions and  
also explores how matter perturbations 
evolve during different cosmological epochs defined by each
critical point.
In our chosen model of DE-DM interaction, 
the coupling parameter $\alpha$ plays a pivotal 
role in  distinguishing the nature of growth rates compared to the scenario 
without matter-quintessence coupling. 
The inclusion of the interaction is found to somewhat prolong 
the occurrence of perturbation growth into the late-time era.\\

The article has been structured in the following way. In Sec.\ \ref{sec:IDS}, we present the field equations for a general interacting field-fluid scenario, encompassing both the equations describing background evolution and those determining the evolution of linear matter perturbations. In Sec.\ \ref{sec:DSA}, we outline the construction of the dynamical system and introduce various relevant cosmological quantities, expressing them in terms of the dynamical variables. In Sec.\ \ref{sec:ROA}, a brief overview of the results obtained from the dynamical stability analysis is provided. We illustrate trajectories in phase space and delineate the evolutionary dynamics of the interacting field-fluid system. Finally, in Sec.\ \ref{con}, we present concluding remarks based on our findings.

\section{Theoretical framework: Evolution at the level of background and perturbations in an  interacting DE-DM scenario }
\label{sec:IDS}
\textbf{Evolution at Background level:} At the outset, we begin with a concise overview of the theoretical framework that delineates the interaction between quintessence dark energy and dark matter in a flat FLRW spacetime background. This FLRW background is characterized by the line-element: 
\begin{eqnarray}
ds^2 &=& \bar{g}_{\mu\nu}dx^\mu dx^\nu = -dt^2 + a^2(t) \left[  \delta_{ij} dx^i dx^j \right] 
\label{eq:metric}
\end{eqnarray}
where $a(t)$ is the scale factor, $t$ denotes cosmic time and
$x^i$'s represent spatial coordinates. 
$\bar{g}_{\mu\nu}$ is the
 unperturbed spacetime metric. 
In the framework of general relativity, the total action governing the universe's dynamics, incorporating dark matter and a dark energy (DE) component driven by a quintessence scalar field $\phi$ that  is minimally coupled
to gravity is given by  \cite{Khyllep:2021wjd}
\begin{eqnarray}
S &=& \int d^4x\sqrt{-g}\left[\frac{R}{2 \kappa^2}+\mathcal{L}_{\phi}+\mathcal{L}_{\rm m}\right] \,,
\label{eq:action_pertds}
\end{eqnarray}
where, $\kappa^2 = 8\pi G$ ($G=$ Gravitational constant), $g$ is the determinant of the metric $\bar{g}_{\mu\nu}$, $R$ denotes the Ricci scalar.
$\mathcal{L}_{\rm m}$  denotes the Lagrangian governing the behaviour 
of the (non-relativistic) dark matter dust,   conceptualized as a perfect fluid characterized by an energy density $\bar{\rho}_{\rm m}$ and a pressure $\bar{p}_{\rm m} = 0$.
$\mathcal{L}_{\phi}$ represents the Lagrangian   of the quintessence scalar field 
$\phi$, responsible for driving dark energy dynamics with minimal coupling to gravity at the background level and is given by \cite{Khyllep:2021wjd}
\begin{eqnarray}
\mathcal{L}_\phi 
&=& 
\frac{1}{2} \bar{g}^{\mu\nu}\partial_\mu \phi \partial_\nu \phi +  V(\phi)\,,
\label{eq:lagphi}
\end{eqnarray}
where $V(\phi)$ is the quintessence potential. Variation of the action with
respect to the metric gives Einstein's field equations 
$R_{\mu\nu}-\frac{1}{2}\bar{g}_{\mu \nu}R = \kappa^2 (\bar{T}_{\mu\nu}^{(\phi)}+\bar{T}_{\mu\nu}^{(\rm m)}) \equiv  \kappa^2 \bar{T}_{\mu\nu}$, where 
 $\bar{T}_{\mu\nu}^{(\rm m)}$ and $\bar{T}_{\mu\nu}^{(\phi)}$
respectively represent the energy momentum tensors of individual DM and DE
sectors and $\bar{T}_{\mu\nu}$ is the energy momentum tensor  of
the total dark fluid (DM+DE) at the background level.
For a homogeneous field $\phi \equiv \phi(t)$,
the energy momentum tensor 
\begin{eqnarray}
\bar{T}_{\mu\nu}^{(\phi)}  =
\partial_\mu \phi \partial_\nu \phi- \bar{g}_{\mu\nu} 
\left[\frac{1}{2}\bar{g}^{\alpha\beta}\partial_\alpha \phi \partial_\beta \phi 
+V(\phi)\right]
= \bar{p}_\phi \bar{g}_{\mu\nu}+(\bar{\rho}_\phi+\bar{p}_\phi)\bar{u}_\mu \bar{u}_\nu\, 
\label{eq:tenphi}
\end{eqnarray}
 mimics  that of an ideal fluid with pressure $\bar{p}_\phi$ and 
energy density $\bar{\rho}_\phi$ given as
\begin{eqnarray}
 \bar{p}_\phi= \frac{1}{2}\dot{\phi}^2 - V(\phi)\,,\quad
\bar{\rho}_\phi = \frac{1}{2}\dot{\phi}^2 + V(\phi)
\label{eq:prhophi}
\end{eqnarray}
On the other hand, the background of the energy-momentum tensor 
$\bar{T}_{\mu\nu}^{(\rm m)}$ for DM dust 
considered as ideal fluid, is given by
\begin{eqnarray}
\bar{T}_{\mu\nu}^{(\rm m)}=\bar{p}_{\rm m} \bar{g}_{\mu\nu}+(\bar{\rho}_{\rm m}+\bar{p}_{\rm m})\bar{u}_\mu \bar{u}_\nu\,,\quad \mbox{with } \bar{p}_{\rm m} = 0.
\label{eq:prhom}
\end{eqnarray}
At the background level, the energy-momentum tensor of the flat universe, encompassing both  DE  and  DM, treated as perfect fluids, can be expressed as  
\begin{eqnarray}
\bar{T}_{\mu\nu} &=& \bar{T}_{\mu\nu}^{(\rm m)} + \bar{T}_{\mu\nu}^{(\phi)}
= \bar{p}\bar{g}_{\mu\nu} + (\bar{p}+\bar{\rho})\bar{u}_\mu \bar{u}_\nu\,,
\label{eq:Ttot}
\end{eqnarray}
where we used $\bar{p} = \bar{p}_\phi$ and 
$\bar{\rho} = \bar{\rho}_{\rm m} + \bar{\rho}_\phi$ which represent 
 pressure and energy density of the total
dark fluid and $\bar{u}_\mu$ is the (unperturbed) 4-velocity
satisfying $\bar{g}_{\mu\nu} \bar{u}^\mu \bar{u}^\nu = -1$.
The overall energy-momentum tensor  $\bar{T}_{\mu\nu}$
adheres to the conservation equation $\nabla_\nu \bar{T}^{\mu\nu} = 0$,
yielding the continuity equation expressed in terms of 
$\bar{\rho}$ and $\bar{p}$ as 
\begin{eqnarray}
\dot{\bar{\rho}} + 3H(\bar{p} + \bar{\rho}) &=& 0\,,
\label{eq:cont}
\end{eqnarray}
where $H \equiv \dot{a}/a$ is the Hubble parameter.
Under the decomposition 
$\bar{T}_{\mu\nu} = \bar{T}_{\mu\nu}^{(\rm m)} + \bar{T}_{\mu\nu}^{(\phi)}$,
the conservation equation 
$\nabla^\mu \bar{T}_{\mu\nu} =   0$ leads to the 
decomposed form of 
continuity equation (\ref{eq:cont}) at the background level
given as
\begin{eqnarray}
\Big{[}\dot{\bar{\rho}}_{\rm m} + 3H  \bar{\rho}_{\rm m}   \Big{]}
+ \Big{[}\dot{\bar{\rho}}_{\phi}+ 3H(\bar{p}_\phi + \bar{\rho}_\phi )\Big{]} &=& 0\,,
\label{eq:cont1}
\end{eqnarray}

The incorporation of interaction between dark matter and dark energy can be executed either by introducing it at the Lagrangian level, guided by field theoretic considerations, or by introducing a non-zero source term into the continuity equation for each interacting sector at a phenomenological level.
We choose the second approach to incorporate interactions at the background level,  through the assertion
$\nabla^\mu \bar{T}_{\mu\nu}^{(\rm m)} = -  \nabla^\mu \bar{T}_{\mu\nu}^{(\phi)}
= - \bar{Q}_\nu \neq 0$ amounting to  writing \cite{Khyllep:2021wjd},
\begin{eqnarray}
\Big{[}\dot{\bar{\rho}}_{\rm m} + 3H  \bar{\rho}_{\rm m}   \Big{]}
= - \bar{Q}_0 =  - \Big{[}\dot{\bar{\rho}}_{\phi}+ 3H(\bar{p}_\phi + \bar{\rho}_\phi )\Big{]}  \,,
\label{eq:cont2}
\end{eqnarray}
Introduction of $ \bar{Q}_0 \neq 0$ in Eq.\ (\ref{eq:cont2})
denotes the presence of a source term in the continuity 
equation for individual sectors
keeping intact the  overall validity of the continuity equation (\ref{eq:cont1}) of the total dark fluid. 
Its magnitude serves as a 
 measure of the rate of energy transfer between the two interacting tensors. Note that,
$\bar{Q}_0>0$ indicates the direction of energy flow from the DM  to  DE  sector,
 while $\bar{Q}_0<0$ 
 signifies energy flow in the opposite direction.
\\

The evolution of the universe is governed by the Friedmann equations
\begin{eqnarray}
	3H^2 &=&\kappa^2 \left(\bar{\rho}_{\rm m}+\bar{\rho}_\phi\right)\,, \label{eq:frde_pertds}\\
	2\dot{H}+3H^2 &= &-\kappa^2   \bar{p}_\phi \,,\label{eq:rce_pertds} 
\end{eqnarray}
Defining the critical density parameters for the two dark sectors as
\begin{eqnarray}
	\Omega_\phi 
	&\equiv&\frac{\kappa^2\bar{\rho}_\phi}{3H^2}=\frac{\kappa^2\dot{\phi}^2}{6H^2}+\frac{
		\kappa^2 V}{3H^2} \,, \label{eq:Omp_pertds}\\
	\Omega_{\rm m} &\equiv& \frac{\kappa^2\bar{\rho}_{\rm m}}{3 H^2}\,, \label{eq:Omm_pertds} 
\end{eqnarray}
Eq.\ (\ref{eq:frde_pertds}) assumes the form
\begin{eqnarray} 
\Omega_{\rm m}+\Omega_\phi=1\,. \label{eq:Om_reln_pertds}
\end{eqnarray}
At the background level, the deceleration parameter $\rm q$ and the total equation-of-state (EoS) parameter 
$\omega_{\rm tot}$ 
are given by
\begin{eqnarray}
\rm q \equiv - \frac{\ddot{a}a}{\dot{a}^2}  \quad  
\mbox{and} \quad
\omega_{\rm tot} =   \frac{\bar{p}_\phi}{ \bar{\rho}_{\rm m} + \bar{\rho}_\phi}
= \frac{2q-1}{3} \cdot 
\end{eqnarray}
The accelerating phase of the universe is achieved for  $\rm q < 0$ or $\omega_{\rm tot} < -\frac13$.\\

\textbf{Evolution at the level of linear perturbations:} We
focus on the
  scalar perturbations to the metric that develop during the era of structure formation
and we express the corresponding perturbed metric in a conformal (Newtonian) gauge
through the line element as \cite{Marcondes:2016reb,Khyllep:2021wjd, Dutta:2017wfd}

\begin{eqnarray}
ds^2 &=& a^2(\tau)\big{[} - (1+2\Psi) d\tau^2 + (1 - 2\Phi)\delta_{ij}dx^idx^j\Big{]}
\label{eq:pert}
\end{eqnarray}
where $\tau$ with  $d \tau \equiv \frac{dt}{a}$ , represents the conformal time
and  $\Psi(x)$ and $\Phi(x)$ ($|\Psi|,|\Phi| \ll 1$) are   functions  parametrising the scalar metric perturbations in
conformal gauge.
We ignore the anisotropic stress for the late time phase of evolution
implying $\Psi = \Phi$  and write the perturbed metric 
$g_{\mu\nu} = \bar{g}_{\mu\nu} + h_{\mu\nu}$ with perturbation $h_{\mu\nu}$
having the diagonal form $h_{\mu\nu} = -2a^2\Phi\delta_{\mu\nu}$. 
 In presence of perturbations, the matter density and 4-velocity   
 are respectively decomposed 
 as $\rho_{\rm m}  = \bar{\rho}_{\rm m} + \delta\rho_{\rm m}$, $u_\mu = \bar{u}_\mu + \delta u_\mu$,
 and the source term $Q_\nu$,
 signifying presence of interactions between DE and DM sectors
 at background level,
 is perturbed to $Q_\nu = \bar{Q}_\nu + \delta Q_\nu$.
 As the growth of structure takes place at spatial scales much smaller than the horizon scales, we focus  solely on perturbations within the matter sector,  disregarding
quintessence field perturbations in the dark energy sector.
We, therefore, write the perturbed energy momentum tensor for the DM sector as
$(  T_{\rm m} )^\mu_{~\nu}
= ({\bar T_{\rm m}})^\mu_{~\nu} + 
{ {(\delta T_{\rm m})}^\mu}_\nu $ 
with
$\delta{ T^0}_0 = -\delta \rho_{\rm m}$,
 $\delta{ T^i}_0 = -a^{-1}(\bar{\rho}_{\rm m})\delta u^i$. 
Employing these perturbations in conjunction with metric perturbations $h_{\mu\nu}$, the evolution equation for perturbations is articulated through the following two equations \cite{Marcondes:2016reb, Khyllep:2021wjd}:
\begin{eqnarray}
-\delta'_{\rm m} + \frac{\bar{Q}_0}{\bar{\rho}_{\rm m}}\delta_{\rm m}- \theta + 3\Phi' &=& \frac{\delta Q_0}{\bar{\rho}_{\rm m}}
\label{eq:evol1}\\
\theta' + \left[\mathcal{H} -
\frac{ \bar{Q}_0}{\bar{\rho}_{\rm m}} \right]\theta
- k^2 \Phi &=& \frac{ik^i\delta Q_i}{\bar{\rho}_{\rm m}}
\label{eq:evol2}
\end{eqnarray}
where $\delta_{\rm m} \equiv \delta \rho_{\rm m}/\bar{\rho}_{\rm m}$
represents the  DM energy density contrast
and  $\theta \equiv a^{-1} ik^j\delta u_j$ denotes the
divergence of the velocity perturbations ($\delta u_j$)
 in the Fourier space,
 and $k^j$ represents the components of the wave vector in Fourier space.
 Here the symbol $^\prime$ denotes differentiation with 
respect to conformal time $\tau$ and $\mathcal{H} = a'/a$. 
We employed $\delta u_0 = -a\Phi$ based on the consideration of the equation $\bar{g}_{\mu\nu}\bar{u}^\mu \bar{u}^\nu = -1$ and also neglected any pressure-perturbation to the  dark matter  dust,
which is    considered to be 
presseureless, at the background level $(\bar{p}_{\rm m}, \delta p_{\rm m} = 0)$. The wave number $k$, representing the scale of perturbations in the universe, is inversely related to the physical size of structures. For scales much smaller than the Hubble horizon ($k \gg \cal{H}$), $k$ typically ranges from $0.01 \, h \, \text{Mpc}^{-1}$ (supercluster scales) to $1 \, h \, \text{Mpc}^{-1}$ (galaxy scales). Our study examines the growth of matter density perturbations in this regime, where gravitational effects dominate, and dark energy perturbations are negligible and the   expression of Poisson equation in Fourier space 
    takes the form \cite{Marcondes:2016reb,Khyllep:2021wjd}: 
\begin{eqnarray}
k^2\Phi &=& - \frac{3}{2} \mathcal{H}^2 \Omega_{\rm m} \delta_{\rm m}\,.
\label{eq:evol3}
\end{eqnarray}
where, $\Omega_{\rm m} =  \bar{\rho}_{\rm m} / \bar{\rho}_{\rm criticial} $.  
Utilizing Eqs.\ (\ref{eq:evol1}, \ref{eq:evol2}, \ref{eq:evol3}) 
and assuming a negligible time variation of the gravitational 
potential $\Phi$, we can represent the evolution governed by the 
two first-order equations (\ref{eq:evol1}) and (\ref{eq:evol2}), in 
terms of a single second-order differential equation for the DM
density contrast as \cite{Marcondes:2016reb}: 
\begin{eqnarray}
\delta''_{\rm m} - \left( \mathcal{Q} - \mathcal{K} \right) \delta'_{\rm m} - \left( \frac{3}{2} \mathcal{H}^2 \Omega_{\rm m} +  \mathcal{Q}' + \mathcal{K} \mathcal{Q} \right) \delta_{\rm m} =
- \frac{i k^i \delta Q_i}{\bar\rho_{\rm m}},
\label{eq:fullSOE}
\end{eqnarray}
where $\mathcal{Q} = \frac{\bar{Q}_0}{\bar{\rho}_{\rm m}} 
- \frac{(\delta Q)}{\bar{\rho}_{\rm m}\delta_{\rm m}}$ and $\mathcal{K} = \mathcal{H} -  \frac{\bar{Q}_0}{\bar{\rho}_{\rm m}}$.
Note that, in the absence of any interaction,   which implies $\bar{Q}_0, \delta Q = 0$, and $\mathcal{K} = \mathcal{H}$, 
Eq.\ (\ref{eq:fullSOE}) reduces to 
$\delta''_{\rm m} + \mathcal{H}\delta'_{\rm m} - 
\frac{3}{2}\mathcal{H}^2 \Omega_{\rm m}\delta_{\rm m} = 0$, 
or alternatively, to $\ddot{\delta}_{\rm m} + 2H\dot{\delta}_{\rm m} - \frac{3}{2}H^2 \Omega_{\rm m}\delta_{\rm m} = 0$, when the involved quantities are expressed in terms of cosmic time $t$ instead of conformal time $\tau$.

\section{Dynamical stability analysis of the interacting system}
\label{sec:DSA}

 The equations that govern the evolution of the universe, accounting for both background and perturbations, are inherently complex, presenting considerable challenges in deriving their analytical or numerical solutions.
To explore the characteristics of cosmic evolution, we utilize dynamical systems analysis,
 which allows us to bypass the requirement for direct analytical or numerical computations.
  In this section, we outline the formulation of cosmic evolution through a set of autonomous equations expressed in terms of dimensionless dynamical parameters. These parameters are appropriately  defined 
  in terms of various cosmological quantities  pertinent to the considered scenario.
 As previously mentioned, we treat DM as pressureless dust, implying an equation of state $\omega_{\rm m} = 0$. We adopt an exponential form for the quintessence potential  $V (\phi) = V_0 e^{-\lambda \kappa \phi}$, which is responsible for driving the dynamics of the DE component, where $V_0$ is a positive constant and $\lambda$ is a dimensionless parameter  which can be both
 positive and negative as revealed from our analysis. 
 We adopt a specific form of the source term, 
 $Q_0 = \alpha \bar{\rho}_{\rm m} (H + \kappa \dot{\phi})$, to explore the evolutionary dynamics of the universe with its DM density perturbations.  
The parameter $\alpha$ quantifies the strength of the DE-DM coupling, with $\alpha>0$ indicating the direction of energy flow from DM to DE, and the opposite direction for $\alpha<0$.

 The evolution of density perturbations in a universe featuring dark energy-dark matter interactions, 
 as characterized by the chosen form of $Q_0$, then follows from Eq. (\ref{eq:fullSOE}). 
 By transforming the independent variable from conformal time $\tau$ to cosmic time $t$, 
 the resulting equation takes the form:
\begin{eqnarray}
\ddot{\delta}_{\rm m} + \dot{\delta}_{\rm m} \left[2H - \alpha(H + \kappa \dot{\phi})\right] - \frac32 H^2 \Omega_{\rm m} \delta_{\rm m} = 0
\label{eq:D1}
\end{eqnarray}
where a single (double) dot(s) over any symbol denotes its first (second) order
derivative with respect to  $t$. 
It is important to highlight that   we have  disregarded perturbations 
to the quintessence dark energy field $\phi$ at scales smaller than the horizon, 
a reasonable assumption
that holds during the era of growth of matter perturbations.
However, by utilizing Eqs.\ (\ref{eq:prhophi}) and (\ref{eq:cont2}), we can express $\dot{\phi}$ as $\sqrt{(\bar{\rho}_\phi + \bar{p}_\phi)}$. Consequently, the 
(unperturbed) energy density of the quintessence dark energy component enters the evolution equation, Eq.\ (\ref{eq:D1}), through the $\dot{\phi}$ term, thereby contributing to the development of matter perturbations. 
Due to the influence of $\bar{\rho}_\phi$ on the evolution of dark matter density perturbations within the framework of the considered model of interactions, the model is potentially equipped to address the cosmic coincidence problem, as revealed in the subsequent analysis discussed in the latter part of the article. 
By substituting Eq.\ (\ref{eq:prhophi}) into the continuity  Eq.\ (\ref{eq:cont2}) for the dark energy sector, we can write the equation of motion for the quintessence field $\phi$ in the presence of interactions as 
\begin{eqnarray}
\ddot{\phi}+3H\dot{\phi}+\frac{dV}{d\phi} = - \frac{\bar Q_0}{\dot{\phi}} = - \alpha \bar{\rho}_{\rm m}\left(\frac{H}{\dot{\phi}}+\kappa \right),
\label{eq:D2}
\end{eqnarray}
In order to examine the evolution of perturbations using the dynamical analysis approach, we transform equations \eqref{eq:D1} and \eqref{eq:D2} into a first-order autonomous system of equations. This transformation is achieved by introducing 
the following auxiliary variables as 
\begin{eqnarray}
	x=\frac{\kappa \dot{\phi}}{\sqrt{6} H}, ~~~y=\frac{\kappa \sqrt{V}}{\sqrt{3} H}, 
	~~~u=\frac{d (\ln\delta_{\rm m} )}{d(\ln a)}\,,
\,  \label{eq:var_pertds}
\end{eqnarray} 
and in their terms the evolution equations  
takes the form
\begin{eqnarray}
	x'&=& -\alpha  \left(\frac{1}{2 x}+\sqrt{\frac{3}{2}}\right) \left(-x^2-y^2+1\right)+\frac{3}{2} x \left(x^2-y^2+1\right)-3 x+\sqrt{\frac{3}{2}} \lambda  y^2 \,, \label{eq:x_1_pertds}\\
	y'&=& -\frac{\sqrt{6}}{2} \lambda xy+\frac{3}{2} y (1+x^2-y^2)\,, \label{eq:y_1_pertds}\\
	u'&=& -u (u+2-\alpha-\sqrt{6}\alpha x) +\frac{3}{2} (1-x^2-y^2) +\frac{3}{2} (1+x^2-y^2) 
u\,. \label{eq:u_1_pertds}
\end{eqnarray}
Here, symbol $^\prime$ signifies derivatives with respect to $\ln a$
and in this notation,  
we can express $u$ as $\frac{\delta_{\rm m}^\prime}{\delta_{\rm m}}$.  By introducing the dimensionless dynamical variables $x$, $y$, and $u$, we rewrite eqn. \eqref{eq:D1} as:
\begin{eqnarray}
\delta''_m + \delta'_m \Big{[}2 - \alpha \left(1 + \sqrt{6} \, x\right)\Big{]} - \frac{3}{2} \delta_m \left(1 - x^2 - y^2\right) = 0 \label{eq:deltadprime}
\end{eqnarray}

Note that, the dynamical variables $x$ and $y$, involving $\dot{\phi}$ and $V(\phi)$ respectively (along with $H$) (eq.\ (\ref{eq:var_pertds})), are related to describing the background dynamics of the universe, while $u$ represents the rate of growth of perturbations over the background.  
A positive value of $u$ indicates the growth of inhomogeneities
with time, while a negative value signifies the decay of inhomogeneities.
In terms of these variables, the background cosmological quantities $\Omega_\phi$, $\Omega_{\rm m}$, $\omega_{\rm tot}$, deceleration parameter ($\rm q$), and the coincidence parameter ($\rm r_{\rm mc}$) can be expressed as: 
\begin{eqnarray}
 \Omega_\phi=x^2+y^2\,,\quad
 \Omega_{\rm m}=1-(x^2+y^2)\,,\quad 
	 \omega_{\rm tot}=x^2-y^2\,,\quad 
	  {\rm q} = \frac12 + \frac32 (x^2-y^2) \,.
	\label{eq:def}
\end{eqnarray}
The coincidence parameter, an important quantity investigated in this study to address the issue of cosmic coincidence, can also be represented in terms of the dynamical variables through its definition, $\rm r_{\rm mc} \equiv \frac{\Omega_{\rm m}}{\Omega_{\rm \phi}}$.

\section{Results of the dynamical analysis}
\label{sec:ROA} 

In this section, we present and discuss the results of the comprehensive analysis of the dynamical equations representing cosmic evolution, as outlined in Sec.\ \ref{sec:DSA}, employing dynamical stability techniques.  
We identified critical points in the system by setting the right-hand side of equations \eqref{eq:x_1_pertds}-\eqref{eq:u_1_pertds} to zero and subsequently investigated their stability characteristics by examining the nature of the eigenvalues of the corresponding Jacobian matrix at the critical points. 
From a physical perspective, a stable critical point with $u>0$ indicates unbounded growth of matter perturbations, while one with $u<0$ suggests that matter perturbations will gradually diminish, representing asymptotic stability against such perturbations. 
A stable fixed point with $u=0$ indicates the asymptotic convergence of matter perturbations to a constant value.\\

We have identified a total of 14 real critical points of the system, listed in Tab. \ref{tab:T1}, and labeled as $A_\pm$, $B_\pm$, $C_\pm$, $D_\pm$, $F_\pm$, $F_\pm$, $G$, and $H$ for reference in subsequent discussions.
The stability features of these critical points, along with constraints on the parameters 
$\alpha$ and $\lambda$ to ensure the reality of the critical points (existence conditions), as well as the values of matter density and the grand  EoS  parameter at these points, are listed in Tab. \ref{tab:T2}.
We can recognize various pairs of fixed points within Tab. \ref{tab:T1}, where a specific ($x$, $y$) set of background coordinates corresponds to different perturbation coordinates $u$. This phenomenon is evident in pairs of fixed points such as $(A_{\pm},B_{\pm})$, $(C_{\pm},D_{\pm})$, $(E_{\pm},F_{\pm})$, and $(G,H)$. 
The particular selection of dynamical variables   in Eq. (\ref{eq:var_pertds})  manifests  the character of matter perturbations  - whether they grow, decay, or remain constant - solely through one of the three coordinates of the fixed points, while keeping the other two coordinates $(x, y)$ unchanged.  
A comprehensive analysis of the nature of the critical points, along with stability criteria and other relevant characteristics, is presented below.

\begin{table}[H]
\centering
\tiny
\begin{tabular}{|c|c|c|c|}
\hline
Critical & \multirow{2}{*}{x} &  \multirow{2}{*}{y} &   \multirow{2}{*}{u}  \\
points  &   &  &   \\
\hline
\hline
& & & \\
$A_{\pm}$ & $\pm 1$ & 0 & 0 \\
& & & \\
$B_{\pm}$ & $\pm 1$ & 0 & $\pm \sqrt{6} \alpha +\alpha +1$ \\
& & & \\
$C_{\pm}$ & $-\frac{\alpha +\sqrt{(\alpha -2) \alpha }}{\sqrt{6}}$ & 0 & $\frac{1}{4} \Big{(}-\alpha  \left(\alpha +\sqrt{(\alpha -2) \alpha }-1\right)$\nonumber\\
& & & $\pm \sqrt{\alpha  \left(6 \sqrt{(\alpha -2) \alpha }+\alpha  \left(-2 \sqrt{(\alpha -2) \alpha }+2 \alpha  \left(\alpha +\sqrt{(\alpha -2) \alpha }-2\right)-5\right)+6\right)+25}-1\Big{)}$ \\
& & & \\
$D_{\pm}$ & $\frac{-\alpha +\sqrt{(\alpha -2) \alpha }}{\sqrt{6}}$ & 0 & $\frac{1}{4} \Big{(}-\alpha  \left(\alpha +\sqrt{(\alpha -2) \alpha }-1\right)$\nonumber\\
& & &   $\pm \sqrt{\alpha  \left(-6 \sqrt{(\alpha -2) \alpha }+\alpha  \left(-2 \sqrt{(\alpha -2) \alpha }+2 \alpha  \left(\alpha +\sqrt{(\alpha -2) \alpha }-2\right)-5\right)+6\right)+25}-1\Big{)}$ \\
& & & \\
$E_{\pm}$ & $-\frac{\alpha -3}{\sqrt{6} (\alpha +\lambda )}$ & $\pm \frac{\sqrt{\frac{(\alpha -3)^2}{\alpha +\lambda }+6 \alpha +2 \alpha  \lambda }}{\sqrt{6} \sqrt{\alpha +\lambda }}$ & $\frac{-\lambda +\alpha  (\lambda +2)+\sqrt{(\alpha -5)^2 \lambda ^2-4 (\alpha -5) \alpha  \lambda -4 (\alpha -12) \alpha -72}}{4 (\alpha +\lambda )} $\\
& & & \\
$F_{\pm}$ & $-\frac{\alpha -3}{\sqrt{6} (\alpha +\lambda )}$ & $\pm \frac{\sqrt{\frac{(\alpha -3)^2}{\alpha +\lambda }+6 \alpha +2 \alpha  \lambda }}{\sqrt{6} \sqrt{\alpha +\lambda }}$ & $-\frac{\lambda -\alpha  (\lambda +2)+\sqrt{(\alpha -5)^2 \lambda ^2-4 (\alpha -5) \alpha  \lambda -4 (\alpha -12) \alpha -72}}{4 (\alpha +\lambda )} $\\
& & & \\
$G$ & $\frac{\lambda }{\sqrt{6}}$ & $\sqrt{1-\frac{\lambda ^2}{6}}$  & 0\\
& & & \\
$H$ & $\frac{\lambda }{\sqrt{6}}$ & $\sqrt{1-\frac{\lambda ^2}{6}}$  & $\alpha  \lambda +\alpha +\frac{\lambda ^2}{2}-2$\\
& & & \\
\hline
\end{tabular}
\caption{List of all critical points of the autonomous system.}
\label{tab:T1} 
\end{table}
\begin{table}[H]
\centering
\tiny
\begin{tabular}{|c|c|c|c|c|}
\hline 
Critical & \multirow{2}{*}{ Existence} &  \multirow{2}{*}{Stability} &   \multirow{2}{*}{$\Omega_{\rm m}$} & \multirow{2}{*}{$\omega_{\rm tot}$} \\
points  &   &  &  & \\
\hline
\hline
$A_{\pm}$ & Always & See fig. \ref{fig:F1}(a) & 0 & 1 \\ 
& & & &\\
$B_{\pm}$ & Always  & Not found & 0 & 1 \\ 
& & & &\\
$C_{\pm}$ & $\alpha \leq 0~\&~\alpha \geq 2$ & See fig. \ref{fig:F1}(b) & $1-\frac{1}{6} \left(\alpha +\sqrt{(\alpha -2) \alpha }\right)^2$ &  $\frac{1}{6} \left(\alpha +\sqrt{(\alpha -2) \alpha }\right)^2$\\ 
& & & &\\
$D_{\pm}$ & $\alpha \leq 0~\&~\alpha \geq 2$ & See fig. \ref{fig:F1}(c) & $\frac{1}{3} \alpha  \left(-\alpha +\sqrt{(\alpha -2) \alpha }+1\right)+1$ & $\frac{1}{6} \left(\alpha -\sqrt{(\alpha -2) \alpha }\right)^2$ \\ 
& & & &\\
$E_{\pm}$  & $\lambda \in \mathbb{R}~\& \Big{(}\Big{(}\alpha <5~\& \Big{(}\lambda <\frac{2 \alpha }{\alpha -5}-2 \sqrt{2} \sqrt{\frac{\alpha ^2-6 \alpha +9}{(\alpha -5)^2}}$ & See fig. \ref{fig:F1}(d) & $-\frac{(\alpha -3) \left(\alpha  \lambda +\alpha +\lambda ^2-3\right)}{3 (\alpha +\lambda )^2}$  & $-\frac{\alpha  (\lambda +3)}{3 (\alpha +\lambda )}$\\ 
&  $\mbox{ or~} \lambda >2 \sqrt{2} \sqrt{\frac{\alpha ^2-6 \alpha +9}{(\alpha -5)^2}}+\frac{2 \alpha }{\alpha -5}\Big{)}\Big{)}\Big{)} \&   \alpha =5 $& & &\\
&  $\&   \Big{(}\Big{(}\alpha <5 ~\& \Big{(}\lambda <\frac{2 \alpha }{\alpha -5}-2 \sqrt{2} \sqrt{\frac{\alpha ^2-6 \alpha +9}{(\alpha -5)^2}}$& & &\\
& $\mbox{ or~} \lambda >2 \sqrt{2} \sqrt{\frac{\alpha ^2-6 \alpha +9}{(\alpha -5)^2}}+\frac{2 \alpha }{\alpha -5}\Big{)}\Big{)}\Big{)} $  & & &\\
& &  & &\\
$F_{\pm}$ &  $\lambda \in \mathbb{R}~\& \Big{(}\Big{(}\alpha <5~\& \Big{(}\lambda <\frac{2 \alpha }{\alpha -5}-2 \sqrt{2} \sqrt{\frac{\alpha ^2-6 \alpha +9}{(\alpha -5)^2}}$ & See fig. \ref{fig:F1}(e) & $-\frac{(\alpha -3) \left(\alpha  \lambda +\alpha +\lambda ^2-3\right)}{3 (\alpha +\lambda )^2}$  & $-\frac{\alpha  (\lambda +3)}{3 (\alpha +\lambda )} $\\ 
& $ \mbox{ or~}\lambda >2 \sqrt{2} \sqrt{\frac{\alpha ^2-6 \alpha +9}{(\alpha -5)^2}}+\frac{2 \alpha }{\alpha -5}\Big{)}\Big{)}\Big{)} \&   \alpha =5 $ & & &\\
& $\mbox{~or }  \lambda >2 \sqrt{2} \sqrt{\frac{\alpha ^2-6 \alpha +9}{(\alpha -5)^2}}+\frac{2 \alpha }{\alpha -5}\Big{)}\Big{)}\Big{)} $ & & &\\
& &  & &\\
$G$ &$\lambda^2 \leq 6 $ & See See fig. \ref{fig:F1}(f) & 0& $\frac{1}{3} \left(\lambda ^2-3\right)$\\ 
& & & &\\
$H$ &$\lambda^2 \leq 6 $ & See See fig. \ref{fig:F1}(g) &0 &$\frac{1}{3} \left(\lambda ^2-3\right)$ \\ 
\hline
\end{tabular}
\caption{Existence and stability criteria of the critical points of the autonomous system along with values of $\Omega_{\rm m}$ and $\omega_{\rm tot}$ at the critical points.}
\label{tab:T2} 
\end{table}

\begin{itemize}
 \item Points $\rm A_{\pm}$: The existence and characteristics of the
specific critical points in the dynamical system are independent of 
model parameters. These points give  $\Omega_{\rm m} = 0$,
which correspond to an exclusively  dark energy driven
phase and  consistently display no growth scenario    with $u=0$.
But these points are not favored by observations as
 they possess  a total equation of state ($\omega_{\rm tot.}=1$), reflective of a stiff matter epoch.  
The shaded region in Fig. \ref{fig:F1}(a) depicts the constraints on model parameters ($\alpha$ and $\lambda$) necessary for the points to exhibit stability, without further imposition of physical conditions due to the model parameter independence of $\Omega_{\rm m}$ and $\omega_{\rm tot.}$.

\item Points $\rm B_{\pm}$: 
These points exhibit characteristics similar to the previously mentioned critical points, representing solutions dominated by stiff dark energy and are consequently not preferred by observation. The evolutions of matter perturbations depend on the coupling parameter $\alpha$. At the $B_{+}$ point, a growing mode of evolution is observed for $\alpha > -\frac{1}{\sqrt{6}+1}$, and a decaying mode for $\alpha <-\frac{1}{\sqrt{6}+1}$. Similarly, at the $B_{-}$ point, a growing mode solution is identified for $\alpha <\frac{1}{\sqrt{6}-1}$, and a decaying mode for $\alpha >\frac{1}{\sqrt{6}-1}$. However, no stable node has
 been identified at  these points. 

\begin{figure}[H]
\centering
\begin{tabular}{cccc}
\includegraphics[width=0.32\textwidth]{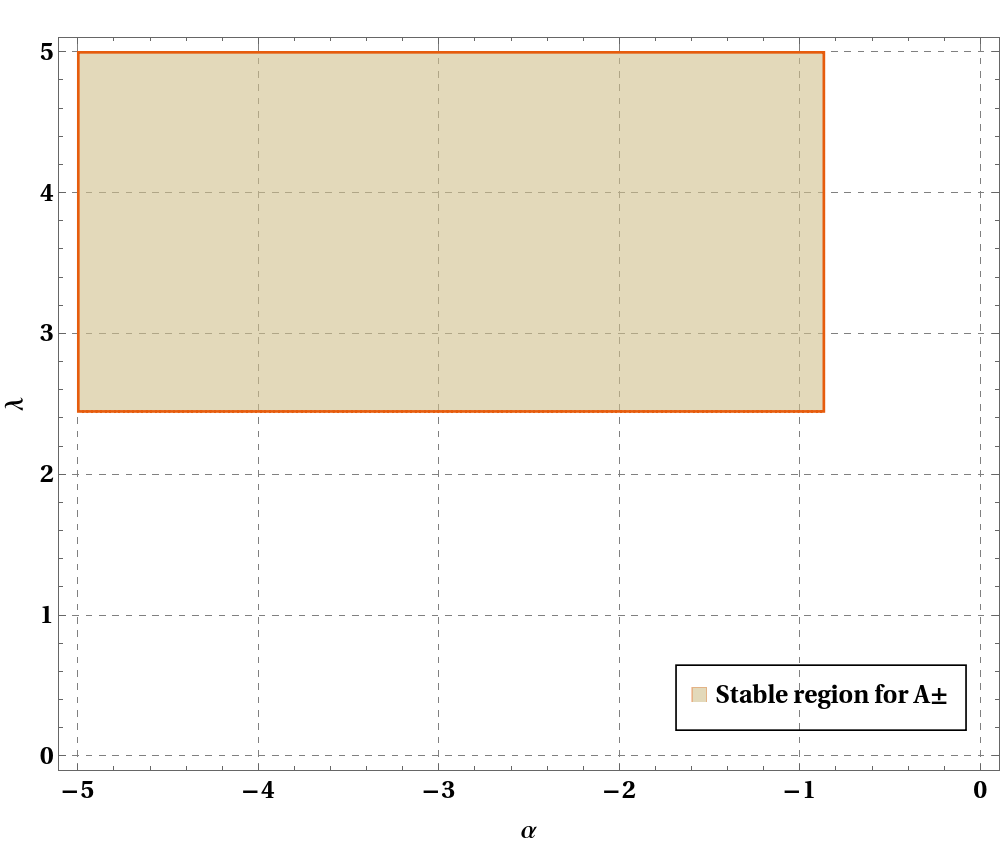} &
\includegraphics[width=0.32\textwidth]{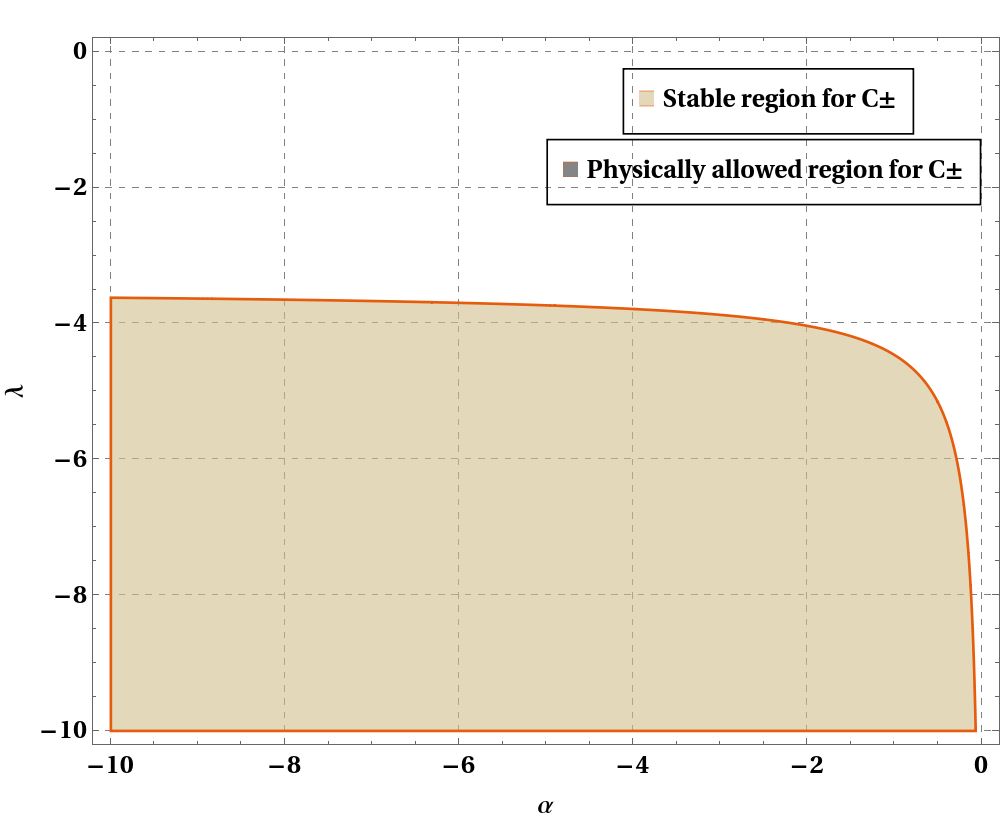} &
\includegraphics[width=0.32\textwidth]{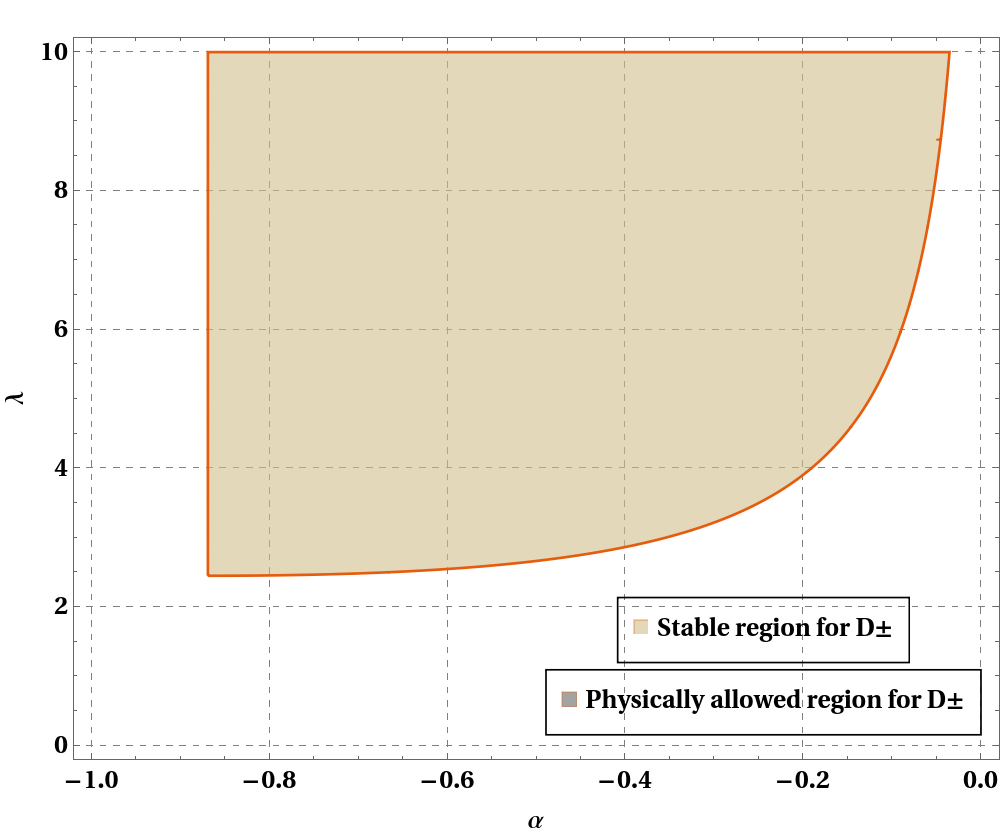} \\
\textbf{(a) Point $\rm A_{\pm}$}  & \textbf{(b) Point $\rm C_{\pm}$} & \textbf{(c) Point $\rm D_{\pm}$}  \\[6pt]
\end{tabular}
\begin{tabular}{cccc}
\includegraphics[width=0.32\textwidth]{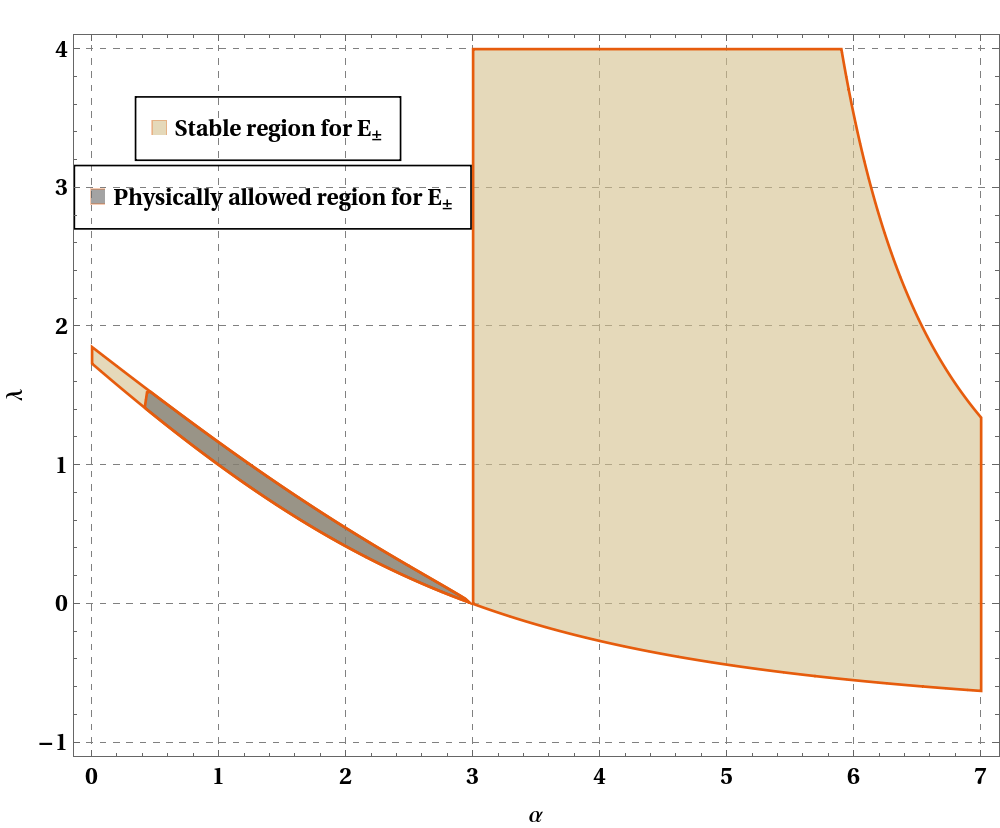} &
\includegraphics[width=0.32\textwidth]{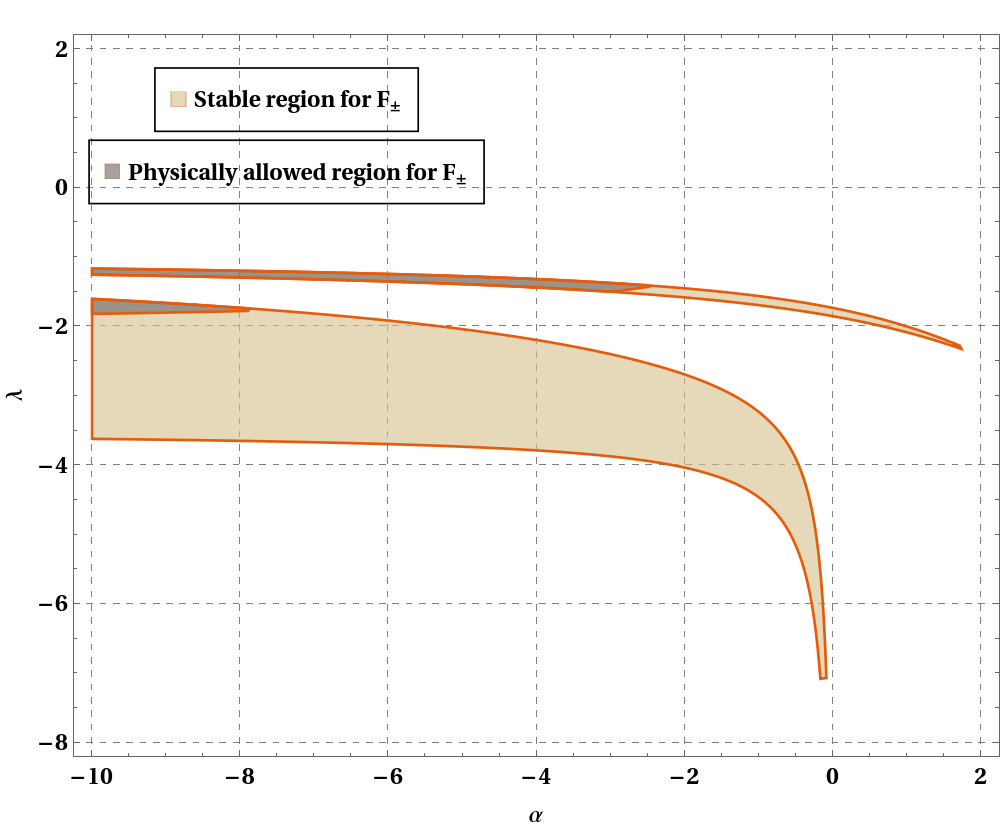} &
\includegraphics[width=0.32\textwidth]{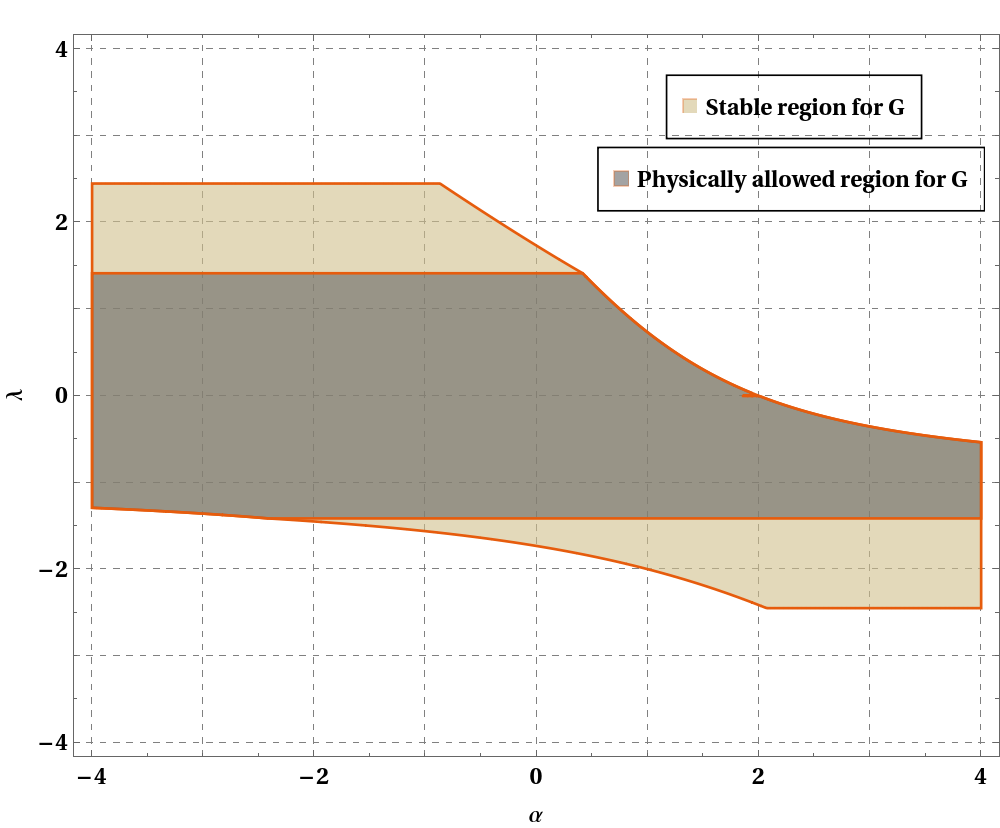} \\
\textbf{(d) Point $\rm E_{\pm}$}  & \textbf{(e) Point $\rm F_{\pm}$} & \textbf{(f) Point $\rm G$} \\[6pt]
\end{tabular}
\begin{tabular}{cccc}
\includegraphics[width=0.32\textwidth]{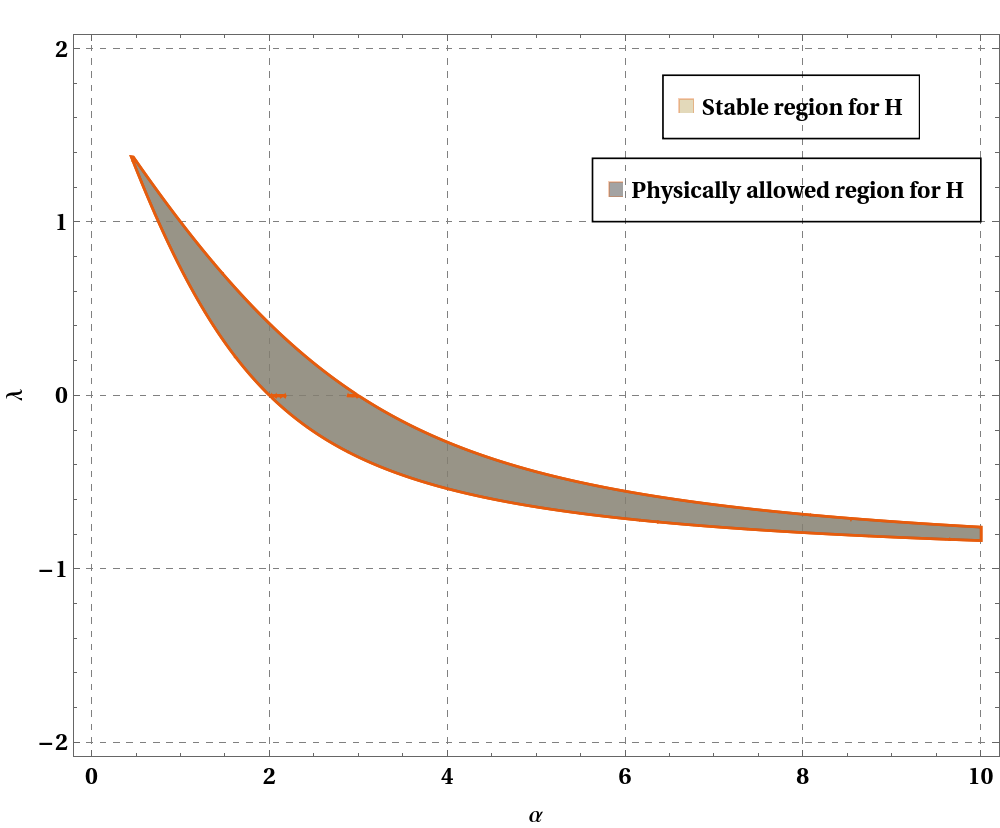} \\
\textbf{(g) Point $\rm H$}  \\[6pt]
\end{tabular}
\caption{The allowed domain in the $\alpha-\lambda$ 
parameter space for which 
the critical points exhibit stability. 
The constraints on parameter space that emerge from imposing conditions $0 < \Omega_{\rm m} < 1$ and  
$-1 < \omega_{\rm tot.} < -\frac13$, which are associated with non-phantom accelerating physically feasible solutions, are also illustrated.} 
\label{fig:F1}
\end{figure}

\item Points $\rm C_{\pm}$:  The characteristics, existence, and cosmological parameters associated with these points depend solely on the coupling parameter $\alpha$. A stable region has been identified for these points, as illustrated in Fig. \ref{fig:F1}(b). However, a stable solution that is physically acceptable, satisfying the critical matter density constraint ($0 < \Omega_{\rm m} < 1$) and the late-time acceleration condition ($-1 < \omega_{\rm tot.} < -\frac13$) simultaneously, has not been identified at these points. Additionally, from Tab. \ref{tab:T1}, it is evident that the growth rate is directly proportional to the coupling parameter, $\alpha$. For $\alpha \leqslant 0$ or $\alpha \geqslant 2$, the growth rate demonstrates a decaying trend.

\item Points $\rm D_{\pm}$:  
The fixed points and the values of cosmological parameters at these points depend   on the coupling parameter $\alpha$. However, their stability characteristics also depend on the quintessence potential parameter ($\lambda$) along with on $\alpha$. Nevertheless, no permissible region that simultaneously satisfies both stability and late-time cosmic acceleration conditions has   been identified (Fig. \ref{fig:F1}(c)). The growth rate is also dependent on the coupling parameter $\alpha$. Furthermore, these points have the potential to exhibit structure formation, and for $\alpha \leqslant 0$ or $\alpha \geqslant 2$, the growth rate depicts a decaying pattern similar to the $\rm C_{\pm}$ points.

\item Points $\rm E_{\pm}$: These critical points are crucial
in the context of late-time stable cosmic accelerating scenario. 
These points, the values of cosmological parameters 
and the  stability characteristics of these points are dependent on
both the model parameters $\alpha$ and $\lambda$.
 A region in the $\alpha-\lambda$ parameter space has been obtained and depicted in
Fig. \ref{fig:F1}(d), which allows for a late-time stable accelerating
solution. The growth rate at these points can also be expressed in terms of both model parameters. These points  exhibit
signatures of growing $(u>0)$ or decaying  $(u<0)$ growth rates
for  parameter regions of $\alpha-\lambda$ depicted in Fig. \ref{fig:F2}(a).
 Negative value of the y-coordinate at the $\rm E_{-}$ point leads to unphysical behavior of the critical point since both the chosen potential form with the constant $V_0$ and the allowed $\lambda$ from our study are positive. Consequently, we have excluded the $\rm E_{-}$ point from our analysis.

\item Points $\rm F_{\pm}$:  
Similar to the previous critical points $E_\pm$, for these points as well, we identify a   region in the $\alpha - \lambda$ parameter space in Fig. \ref{fig:F1}(e), where the points demonstrate stability and also fulfill the physical conditions regarding critical matter density and the grand EoS parameter corresponding to acceleration solutions. These critical points display both increasing and decreasing growth rates, as depicted in Fig. \ref{fig:F2}(b). For similar reasons as  mentioned for the $\rm E_{-}$ point, the $\rm F_{-}$ point, having the same features as $\rm E_{-}$, has also been excluded from our analysis.

\item Point $\rm G$: The characteristics of this specific critical point and the associated cosmological parameter values depend solely on the quintessence potential parameter $\lambda$, without any dependence on the coupling parameter $\alpha$. 
However, its stability characteristic is determined by both $\lambda$ and $\alpha$, as depicted in Fig. \ref{fig:F1}(f). Moreover,  
for this point, we have identified a region in the parameter space where the conditions for stability and physical viability owing to the constraints of $\Omega_{\rm m}$ and late-time acceleration, are simultaneously satisfied. 
This point, which presents a late-time stable accelerating scenario, corresponds uniquely to a  no growth scenario  of perturbations - a feature unique to this point compared to all other identified critical points.

\item Point $\rm H$: 
Similar to the preceding critical point $G$, the characteristics of this specific critical point and the associated cosmological parameter values solely depend on $\lambda$. Additionally, the values of the `background' coordinates $x, y$ for this point are identical to those for point $G$. However, unlike the case of $G$, the `perturbation' coordinate $u$ is not identically zero; rather, it depends on both $\alpha$ and $\lambda$. In Fig. \ref{fig:F1}(g), we have presented the region of parameter space for which the point exhibits stable acceleration and also satisfies other physicality conditions. The point exhibits a dark energy-dominated epoch with a non-phantom type solution for the coupling parameter constraint: $-2 \sqrt{\frac{2}{3}} < \lambda < -\sqrt{2}$ or  $\sqrt{2} < \lambda < 2 \sqrt{\frac{2}{3}}$. In contrast to  no growth scenario seen for point $G$, this point demonstrates a growth rate with $u>0$ or $u<0$, which depends on both the coupling and potential parameters of the system, as depicted in Fig. \ref{fig:F2}(c). 
\end{itemize}
\begin{figure} [H]
\centering
\begin{tabular}{cccc}
\includegraphics[width=0.32\textwidth]{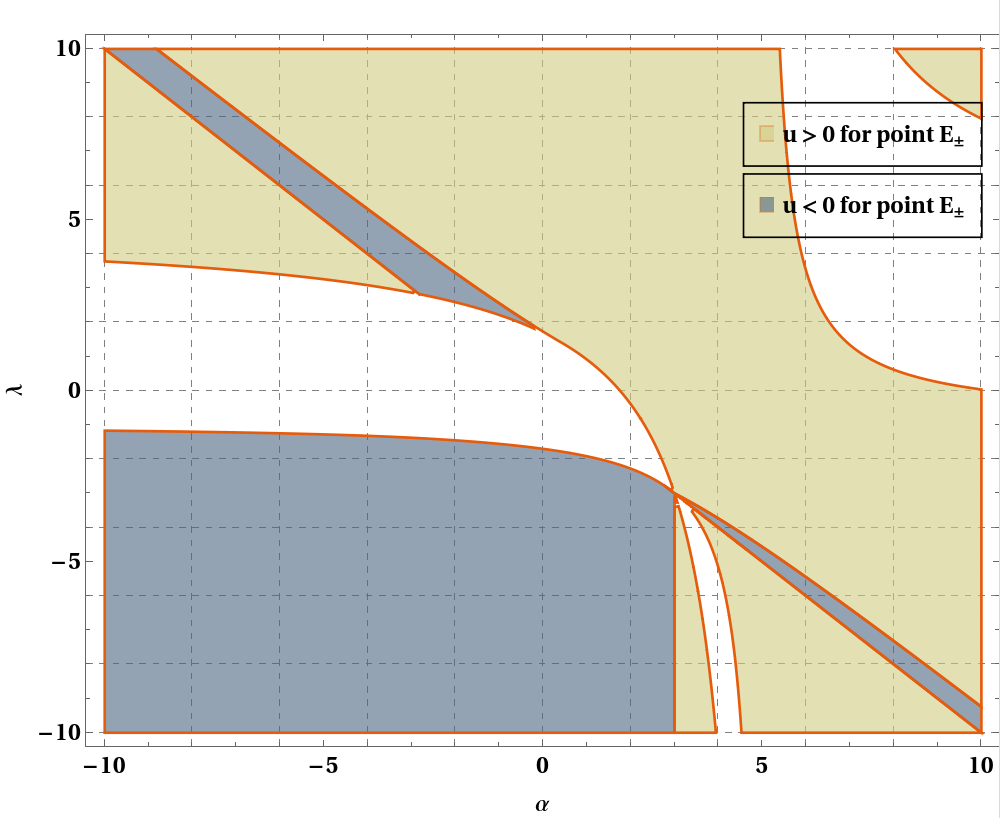} &
\includegraphics[width=0.32\textwidth]{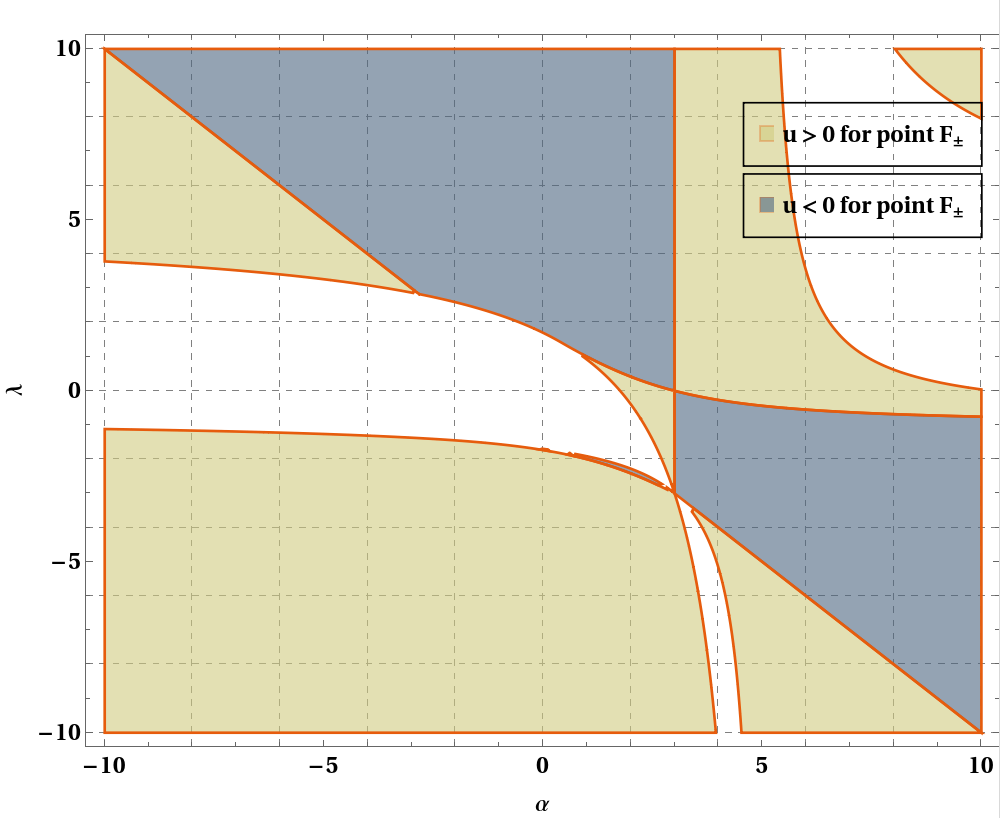} &
\includegraphics[width=0.32\textwidth]{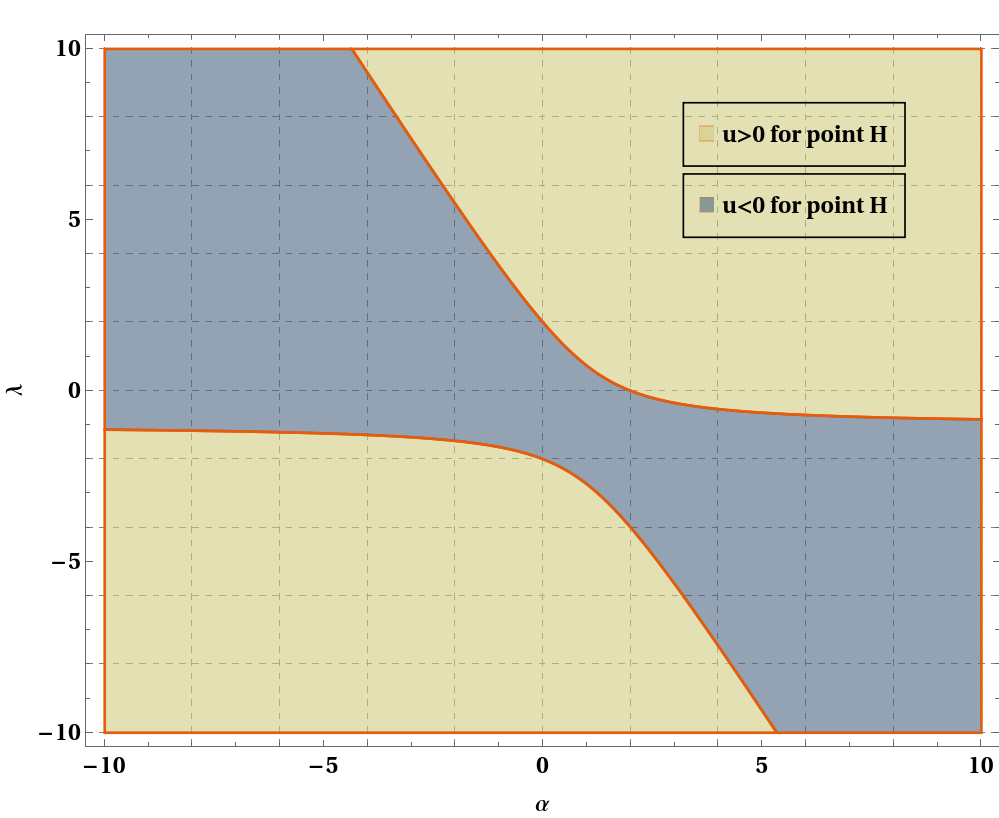} \\
\textbf{(a) Point $\rm E_{\pm}$}  & \textbf{(b) Point $\rm F_{\pm}$} & \textbf{(c) Point $\rm H$}  \\[6pt]
\end{tabular}
\caption{Regions in $\alpha - \lambda$ parameter space
exhibiting growing ($u>0$) or decaying  ($u<0$) type of 
growth rate of matter perturbations corresponding to 
the critical points $E_{\pm}$, $F_{\pm}$ and $H$.}
\label{fig:F2}
\end{figure}

 The ($\alpha-\lambda$) values found within the shaded regions of the parameter space depicted in the subfigures of Fig. \ref{fig:F1}, associated with each fixed point, represent parameter configurations where the fixed point is stable and physically viable. This viability entails adhering to constraints: $0 < \Omega_{\text{m}} < 1$ and $-1 < \omega_{\text{tot}} < -\frac13$, indicating a stable, non-phantom type accelerating solution.\\

\begin{figure}[H]
\centering
\includegraphics[width=0.55\textwidth]{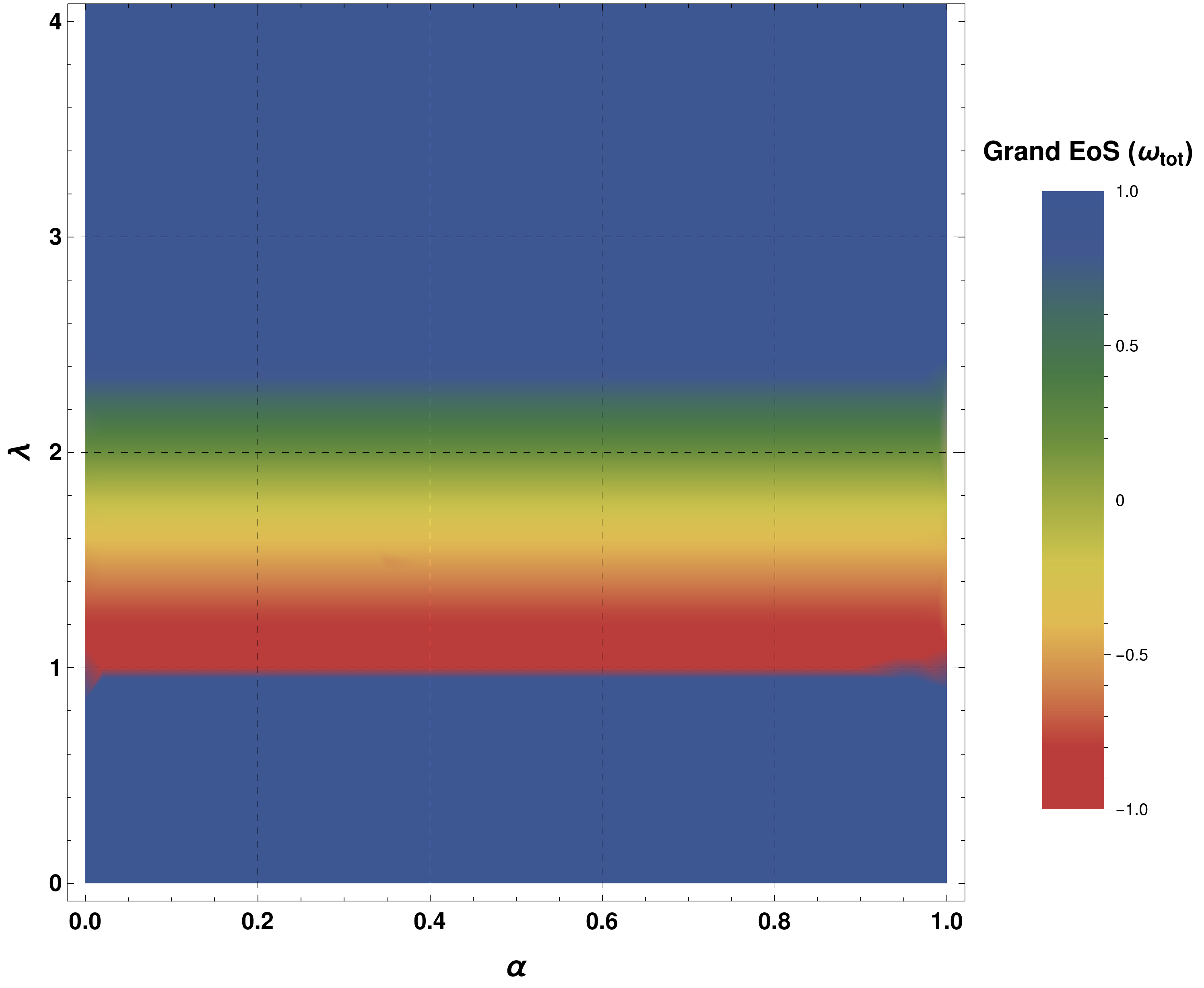}
\caption{Density plots illustrating a variation of grand EoS parameter in $\alpha$-$\lambda$ parameter space.} 
\label{fig:F2a}
\end{figure}

In Fig. \ref{fig:F2a}, we  presented a density plot demonstrating the variation
 of the grand EoS parameter, $\omega_{\rm tot}$ in the ($\alpha$-$\lambda$) plane, 
 while adhering to the constraint $0 < \Omega_{\rm m} < 1$.
 Within the interval $1<\lambda <2.3 $ the  $\omega_{\rm tot}$ deviates from 
 the value of 1. Beyond this interval,  $\omega_{\rm tot}$ maintains a consistent value of 1, indicating a stiff matter phase.  
 Furthermore, within the narrower interval of $1<\lambda<1.4$, the value of $\omega_{\rm tot}$ lies between $-1$ and $-\frac13$,  signifying  accelerated phase of expansion. All the aforementioned ranges of the parameter $\lambda$ are 
 determined by scanning across values of $\alpha$ ranging from 0 to 1.
Since this particular accelerating region is a primary focus of our investigation, we have chosen the benchmark values ($\alpha = 0.01, 1$ and $\lambda=1$) for our subsequent analysis of phase space trajectories and evolutionary dynamics within this interacting dark energy-dark matter system.  We did not find any significant late-time acceleration or viable energy density constraints $(0 < \Omega_m < 1)$ in the region of negative $\alpha$ and $\lambda$, and hence excluded those regions from the   plot in  Fig. \ref{fig:F2a}.  
\\

\begin{figure}[H]
\centering
\begin{subfigure}[b]{0.49\textwidth}
\includegraphics[width=\textwidth]{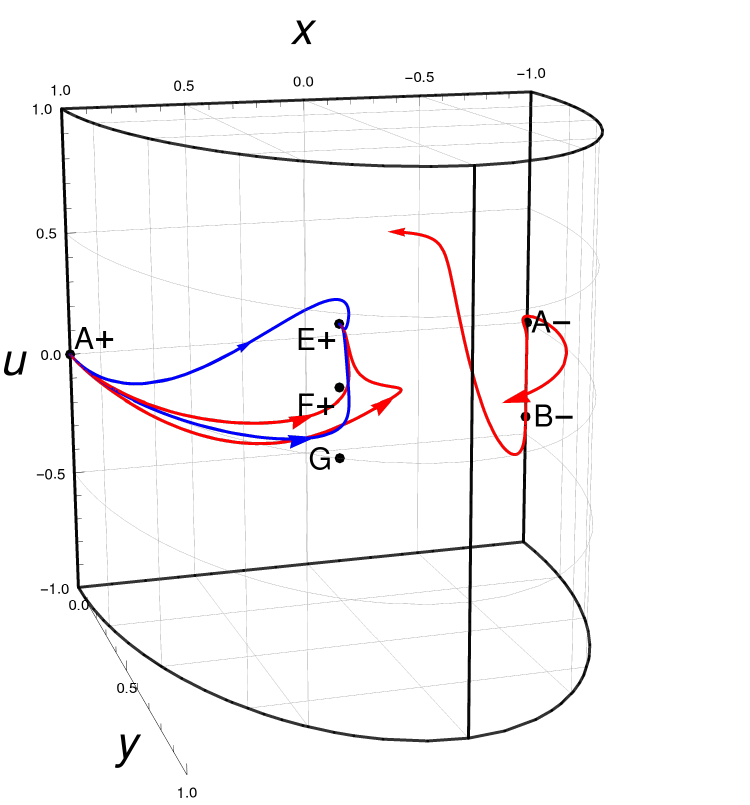}
\end{subfigure}
\begin{subfigure}[b]{0.49\textwidth}
\includegraphics[width=\textwidth]{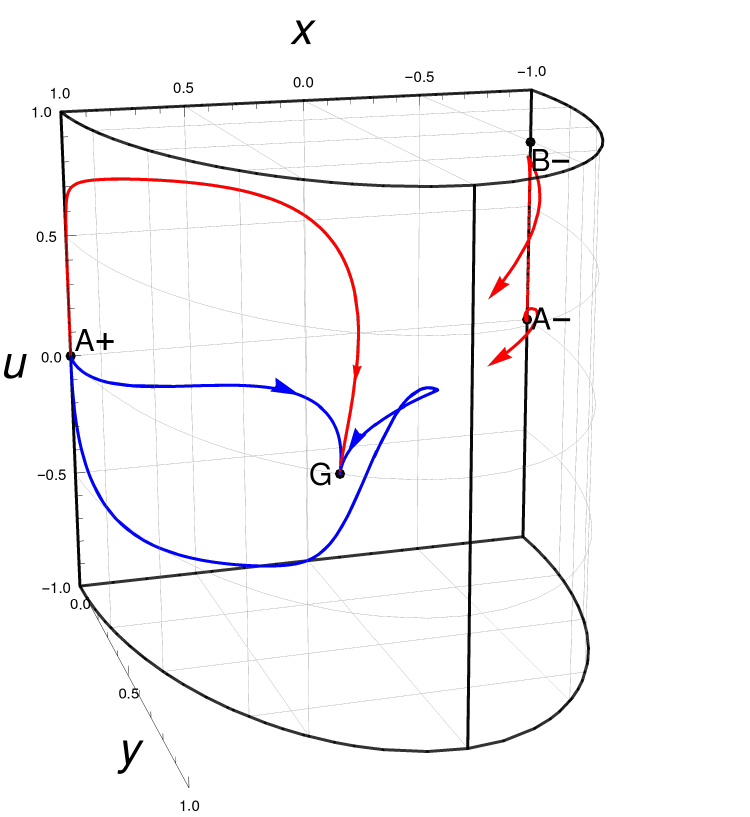}
\end{subfigure}
\caption{Phase space trajectories of the autonomous system,
with  $\alpha=1, \lambda=1$  (left panel) for which
$E_{+}$  behaves as an attractor and with 
 $\alpha=0.01, \lambda=1$ (right panel) for which $G$ serves as an attractor.} 
\label{fig:F3}
\end{figure}

This choice allows us to encompass both low and moderate coupling between the dark energy and pressureless dark matter while maintaining a fixed value for the potential parameter $\lambda (= 1)$. We  illustrated the evolution in terms of the trajectory of the phase point $(x,y,u)$ in three-dimensional space spanned  by these dynamical variables.
 We have depicted the phase trajectories in Fig. \ref{fig:F3} near those critical points that remain real for these parameter choices.  
 The coupling parameter $\alpha$ is found to significantly influence the dynamics around the critical points. Under small coupling conditions ($\alpha=0.01$), critical point $G$ functions as an attractor. Conversely, with moderate coupling ($\alpha=1$), the nature of critical point $G$ shifts from being an attractor to a repeller. Simultaneously, at this specific coupling value, $E_{+}$ transforms into an attractor.  As the set of autonomous equations (\ref{eq:x_1_pertds},\ref{eq:y_1_pertds},\ref{eq:u_1_pertds}) remains invariant under the transformation $y \rightarrow -y$, we only 
 depict in Fig. \ref{fig:F3} the phase-space region corresponding to $y \geqslant 0$. Additionally, the constraint $0<\Omega_{\rm m}<1$ imposes the condition $x^2 + y^2 = 1$. 
 Therefore, the 3-dimensional phase space associated with the system of autonomous equations, 
 which combines background coordinates $x$ and $y$ with perturbation coordinate $u$, 
 corresponds to the set:  
 $\{(x,y,u) \in \mathbb{R}^3 \Big{|} -1\leqslant x \leqslant 1 ; 0 
 \leqslant y \leqslant 1$ and $0 
 \leqslant x^2 + y^2 \leqslant 1\}$.    Considering that all variables, except $u$, have finite bounds, we also investigated the potential presence of critical points at infinity by applying the transformation $u \rightarrow U = \tan^{-1}u$ with $-\frac{\pi}{2} < U < \frac{\pi}{2}$. However, our investigation revealed no existence of critical points at infinity.\\

The left panel of Fig. \ref{fig:F3} displays the phase space trajectories corresponding to the benchmark values $\alpha=1$ and $\lambda=1$. In this scenario, a total of six critical points have been identified. The remaining points either become complex or do not satisfy the physicality condition $0 < \Omega_{\rm m} < 1$. All trajectories are found to converge towards the critical point $E_{+}$ from other points, establishing it as a global attractor. Trajectories depicted in red represent repelling behavior, curving away from saddle or unstable points and heading towards the global attractor. Conversely, trajectories drawn in blue signify attracting behavior, directly converging towards the attractor point.
At these chosen benchmark parameter values, the critical point exhibits dark energy domination with a grand equation of state parameter value of $-0.667$. The nature of the $E_{+}$ point suggests that the interacting DE-DM system can generate a stable, non-phantom type of dark energy-dominated solution, potentially contributing to late-time cosmic acceleration. Since critical point $H$ merges with $E_{+}$ at these particular benchmark values, we have not included point $H$ separately in the phase-space analysis.\\

The right panel of Fig. \ref{fig:F3} illustrates the phase space trajectories corresponding to the benchmark values $\alpha=0.01$ and $\lambda=1$. In this scenario, four critical points have been identified, and the remaining points either become complex or do not satisfy the physical condition. All trajectories converge towards the critical point $G$ from other points, indicating it is a global attractor. Trajectories depicted in red indicate repelling behavior, diverging from saddle or unstable points and moving towards the global attractor. Conversely, blue trajectories signify attracting behavior, directly approaching the attractor point.
At this chosen benchmark value, the critical point becomes dominated by dark energy (DE) with a grand Equation of State (EoS) parameter of $-0.667$. The characteristics of the $G$ point suggest that the interacting field-fluid system has the capability to generate a stable, non-phantom type of dark energy-dominated solution, potentially contributing to late-time cosmic acceleration.

\begin{figure}[H]
\centering
\begin{subfigure}[b]{0.49\textwidth}
\includegraphics[width=\textwidth]{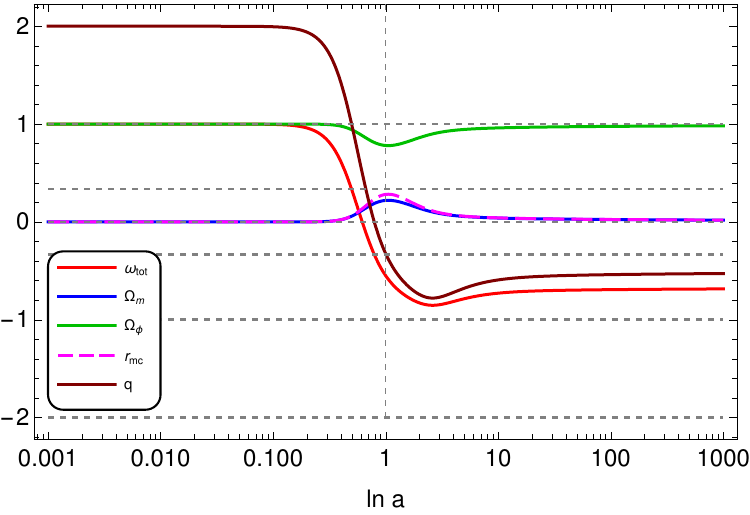}
\end{subfigure}
\begin{subfigure}[b]{0.49\textwidth}
\includegraphics[width=\textwidth]{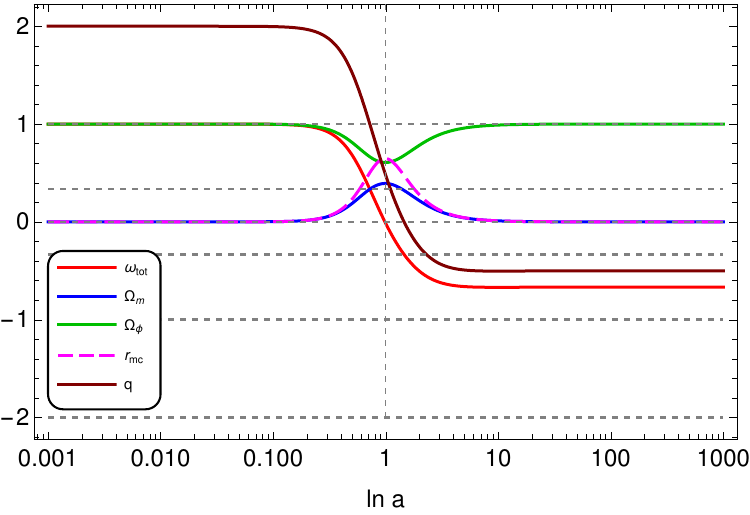}
\end{subfigure}
\caption{Evolution plot for various cosmological parameters
for $\alpha=1, \lambda=1$ (left panel) and
for $\alpha=0.01, \lambda=1$ (right panel).    The initial conditions used to generate the evolution plots of cosmological parameters (\(\omega_{\rm tot}\), \(\Omega_m\), \(\Omega_{\phi}\), \(r_{\rm mc}\), \(q\)) in Fig. 5 are as follows:   For the attractor \(E_+\): \(x(t=0.18) = 0.25\), \(y(t=0.18) = 0.85\), \(u(t=0.18) = 0.6\), and for the attractor \(G\): \(x(t=2.2) = 0.399\), \(y(t=2.2) = 0.91\), \(u(t=2.2) = 0.07.\) 
} 
\label{fig:F4}
\end{figure}

To examine the behavior of various cosmological parameters such as $\omega_{\text{tot}}$, $\Omega_{\text{m}}$, $\Omega_{\phi}$, $r_{\text{mc}}$, and the deceleration parameter $\rm q$ around critical points, we illustrate their evolution profiles with respect to the logarithmic function of the FLRW scale factor in Fig. \ref{fig:F4}. 
We used particular valid initial boundary conditions that enable us to derive solutions for ($x, y, u$), thus furnishing the profiles for the cosmological quantities mentioned above.
  We present these profiles for two sets of benchmark parameter values: $(\alpha=1, \lambda=1)$ in the left panel and $(\alpha=0.01, \lambda=1)$ in the right panel of Fig. \ref{fig:F4}.
We observe that, in both cases, the energy density of the DE sector dominates over the DM sector throughout the evolution, from the early to the late phase.
However, a distinctive feature emerges during the matter-dominated phase. The decrease in the value of the  EoS parameter $\omega_{\text{tot}}$, resulting in its eventual crossing of the zero line, aligns with the attainment of a peak in matter density $\Omega_{\text{m}}$, accompanied by simultaneous occurrence of a trough in the value of the field density parameter $\Omega_{\phi}$.  
As a result, the coincidence parameter $r_{\text{mc}} \equiv \Omega_{\rm m}/\Omega_\phi$ reaches its peak at this epoch, and it gradually decreases during the late-time era, as indicated by the plots.\\

 In this context, it's important to observe that the ratio of matter to curvature density, denoted as $r_{\rm mc} \equiv \frac{\Omega_{\rm m}}{\Omega_{\phi}}$, serves as an indicator of the  dominance of one component over the other.  A value of $r_{\rm mc} = 0$ suggests complete field dominance, while  the value of  $r_{\rm mc}$ of  order  1, suggests a convergence of matter and curvature energy densities in the universe, particularly relevant in the present epoch. The evolution of the parameter $r_{\rm mc}$ reflects the extent of  DE  dominance over  DM  at various stages of evolution. The value of the coupling parameter $\alpha$ exhibits significant implications in this evolutionary process. We observe a greater dominance of DE over DM with higher $\alpha$ values around the present epoch. A higher $\alpha$ indicates a stronger interaction between the two sectors, leading to more energy transfer from DM to DE ($\bar{Q}_0 \propto \alpha$).  
This characteristic is evident from both panels of Fig. \ref{fig:F4}, which correspond to $\alpha = 1$ (left panel) and $\alpha = 0.01$ (right panel), with a constant value of the quintessence potential parameter $\lambda = 1$.  
 Therefore, the issue of cosmic coincidence may also be addressed through the introduction of this form of coupling between the DE and DM sectors.\\

   In order to comprehend the kinematic aspects of cosmic evolution, we have examined the evolution of the deceleration parameter ($\rm q$) for both sets of benchmark values
   of parameters ($\alpha,\lambda$). During the early stages of evolution, the deceleration parameter  remains positive, signifying a decelerating phase. However, at a certain point, the deceleration parameter ($\rm q$) experiences a sharp decline, crosses the zero line, and turns negative, indicating onset of cosmic acceleration, ultimately stabilizing at $-\frac12$. 
   Moreover, the profile of the grand  EoS parameter begins at a positive unity, indicating the stiff matter phase, and transitions through the radiation-dominated epoch where the field density contribution is maximum with negligible dark matter contribution. As the EoS parameter approaches zero, a shift in both energy densities is observed. Eventually, the EoS parameter crosses the $-\frac{1}{3}$ line and stabilizes at $-\frac{2}{3}$, indicating the achievement of a stable accelerating solution for the system. Based on the analysis, we conclude that in both benchmark scenarios, where either $E_{+}$ or $G$ acts as the attractor, the system reaches a stable non-phantom
cosmic accelerating phase within this interacting framework.

\begin{figure}[H]
\centering
\includegraphics[width=0.75\textwidth]{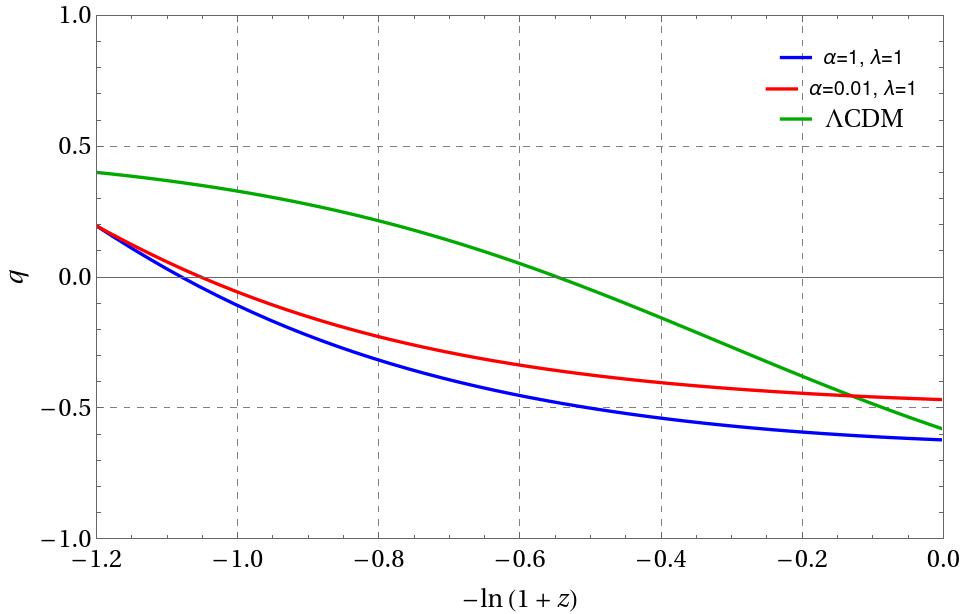}
\caption{ Comparative study of the deceleration parameter with redshift for the interacting model  discussed in the paper and the $\Lambda$-CDM case.}  
\label{fig:FN1}
\end{figure}

In our recent work \cite{Chatterjee:2024rbh}, we analyzed the posterior distributions of $\Lambda$-CDM parameters (\(H_0\), \(\Omega_m^0\), \(M\)) using MCMC samples and a combined dataset (Pantheon, OHD, BAO). This provided 1-$\sigma$ and 3-$\sigma$ uncertainties as:  
\begin{eqnarray}
\Omega^0_{m} = 0.28^{+0.009}_{-0.009}, \quad M = -19.39^{+0.015}_{-0.015}, \quad H_0 = 68.74^{+0.56}_{-0.56} \, \text{km s$^{-1}$ Mpc$^{-1}$}.
\end{eqnarray}
Using these datasets over \(0 \leq z \leq 2.3\) (corresponding to \(t \in [0.23, 1]\), with \(t = 1\) at the present epoch), we plotted the deceleration parameter (\(q = -1 - \frac{\dot{H}}{H^2}\)) for the \(\Lambda\)-CDM model, as shown in Fig.\ \ref{fig:FN1}. For the coupled quintessence scenario, using \(q = \frac{1}{2} + \frac{3}{2}(x^2 - y^2)\), we derived the deceleration parameter’s evolution for \(\alpha = 1, 0.01\) and \(\lambda = 1\). 
All models show a transition from deceleration to acceleration. In the observable redshift range, the deceleration parameter \(q\) approaches \(\sim -0.6\) for both \(\Lambda\)-CDM and the coupled model (\(\alpha = 1, \lambda = 1\)).

\begin{figure}[H]
\centering
\includegraphics[width=0.75\textwidth]{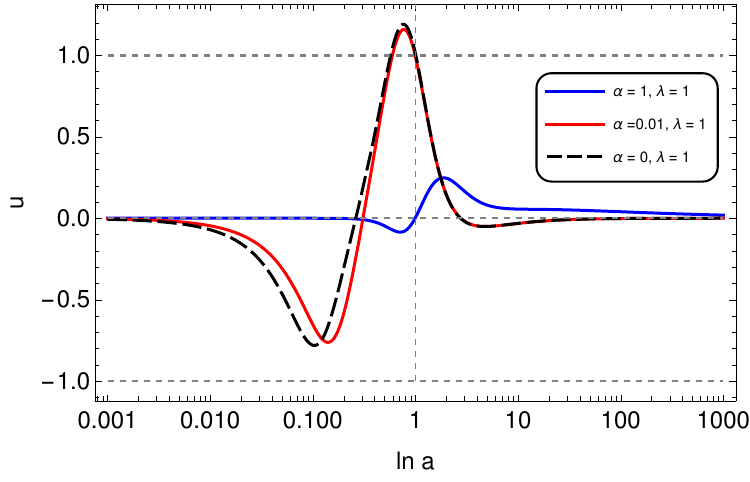}
\caption{Evolution plot of growth rate  ($u$) of matter perturbations for three different values of coupling parameter \textit{viz.}
 $\alpha= 0, 0.01$ and $1$  with fixed $\lambda$ (= 1).
 The initial conditions chosen for generation of the plots are: \(x(t=0.7) = 0.312\), \(y(t=0.7) = 0.865\), \(u(t=0.7) = 0.06\). } 
\label{fig:F5}
\end{figure}

The evolution of the growth rate of matter perturbations captured through the
variable $u$, is shown in Fig.\ \ref{fig:F5}.  We present the trend of $u$ against $\ln a$ for three distinct values of $\alpha$ (0, 0.01, 1), with $\lambda$, held constant at 1. We observe that for $\alpha = 0$ (indicating no DE-DM interactions) and $\alpha = 0.01$ (representing small DE-DM coupling), the growth rate experiences a significant swing  between negative and positive values throughout the evolutionary phase. Notably, the growth rate is high during the matter-dominated phase and gradually diminishes over time. As
the universe  enters the phase of cosmic acceleration, the growth rate
 rapidly converges towards zero. In a scenario characterized by relatively stronger DE-DM interactions ($\alpha = 1$), the amplitude of the growth rate remains significantly suppressed compared to the previous cases (no coupling and weak coupling) throughout the evolutionary era. However, unlike the other two cases, it doesn't sharply fall to zero; instead, it asymptotically approaches zero. 
Therefore, in the presence of a sufficiently moderate interaction between DE-DM sectors, a growth rate of matter perturbations doesn't diminish solely at the onset of the late-time epoch; instead, it can remain significant even in the distant future.\\

Hence, we can precisely distinguish between the matter and dark energy epochs both at the background and perturbation levels. Moreover, the coupling parameter enables us to adjust both background and perturbation behaviors.  
Thus, within suitable parameter ranges, we can achieve the evolutionary dynamics of the universe that are consistent with observations. This includes a matter-dominated phase characterized by substantial growth of matter perturbations, followed by a transition to the stable, non-phantom dark energy-dominated accelerating phase.\\

\begin{figure}[H]
\centering
\begin{subfigure}[b]{0.49\textwidth}
\includegraphics[width=\textwidth]{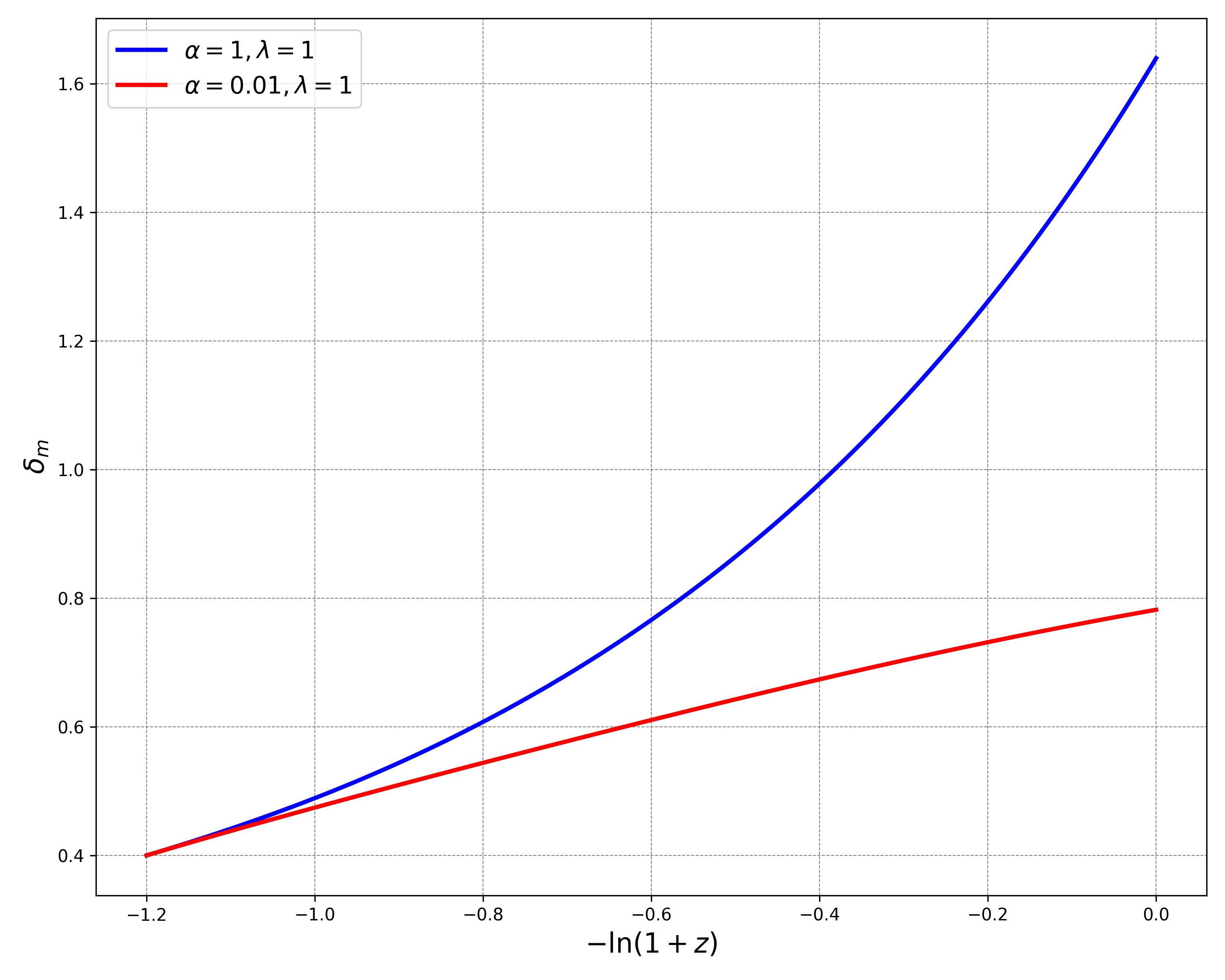}
\end{subfigure}
\begin{subfigure}[b]{0.49\textwidth}
\includegraphics[width=\textwidth]{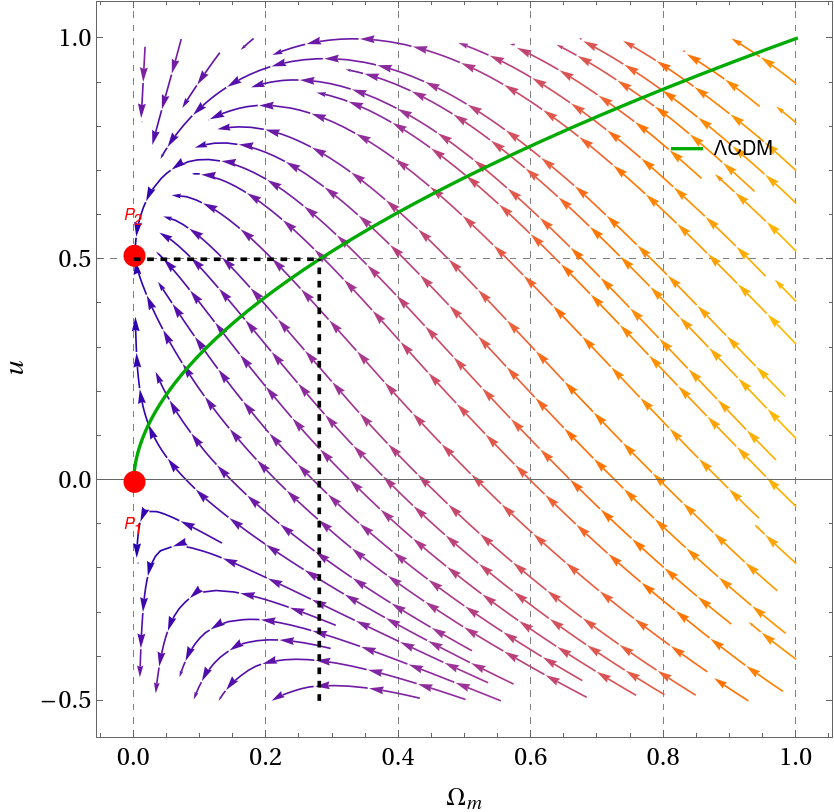}
\end{subfigure}
\caption{Left panel:  Temporal behavior of dark matter density contrast at different values of coupling parameter \textit{viz.}  $\alpha= 1$ and  $0.01$ with fixed $\lambda$ (= 1). Right panel: Phase-space behavior of the growth rate and dark matter energy density for $\alpha = 1$ and $\lambda = 1$. The green line represents the comparison of the growth rate, which follows the parametrization $u = \Omega_m^\gamma$ with $\gamma = \frac{6}{11}$ (for $\Lambda$-CDM    Model). }
\label{fig:FN2}
\end{figure}

Using the equations in Sec.\ \ref{sec:DSA}, we determined the temporal evolution of the dark matter density contrast parameter \( \delta_m \) for two coupling values (\( \alpha = 1, 0.01 \)) with a fixed potential parameter (\( \lambda = 1 \)), corresponding to the stable fixed points of the interacting system. The left panel of Fig.\ \ref{fig:FN2} shows \( \delta_m \) as a function of \( \ln a \sim -\ln(1+z) \), where \( a = \frac{1}{1+z} \) and \( z \) is the redshift, plotted over the range \( 0 < z < 2.3 \) using recent observational datasets. 
Initial conditions for \( x \) and \( y \) were chosen near their stable fixed points, with \( \delta = 0.4 \) and \( \delta' = 0.39 \) to ensure stable evolution. The blue and red lines in the plot correspond to the coupling strengths \( \alpha = 1 \) and \( \alpha = 0.01 \), respectively. The plot highlights that a moderate coupling strength significantly affects the density contrast at late times. 
This effect is further supported by Fig.\ \ref{fig:F5}, which illustrates how the coupling strength influences the growth factor's evolution. For moderate interaction between dark energy and dark matter, the growth rate of matter perturbations does not vanish entirely during the late-time epoch.\\

The growth rate \( u \) is defined as the logarithmic derivative of the matter perturbation with respect to the logarithm of the scale factor: 
$u \equiv \frac{d (\ln \delta_m)}{d (\ln a)}$, 
and is commonly parametrized as: $u = \Omega_m^\gamma$, 
where \( \gamma \) is the growth index. This approximation is effective for most cosmological models without dark energy-dark matter (DE-DM) coupling, with \( \gamma \) varying accordingly. For the \( \Lambda \)CDM model, the growth index is about \( \frac{6}{11} \). However, for the interacting DE-DM model with a source term of the form \( \alpha \bar{\rho}_m (H + \kappa \dot{\phi}) \), expressing the growth index \( \gamma \) becomes impractical. This interaction cannot be described within the standard \( \Omega_{\text{DE}} \) and \( \Omega_{\text{DM}} \) framework, as it introduces a direct coupling that alters the dynamics of both \( \rho_m \) and \( \rho_{\text{DE}} \). This coupling prevents their individual energy densities from being expressed independently, making the parametrization of \( \gamma \) in terms of \( \Omega_{\text{DE}} \) unfeasible. As the evolution of \( \delta_m \) and the growth factor ($u$) becomes intertwined with scalar field dynamics, deriving \( \gamma \) analytically is no longer possible within the standard framework.\\

To explore the effect of coupling strength, particularly for the \( \alpha = 1, \lambda = 1 \) case, we analyzed the dynamical phase-space behavior of the growth rate (\( u \)) and the dark matter energy density parameter (\( \Omega_m \)). We introduced a new autonomous equation for \( \Omega_m \) to study the growth rate and growth index, and compare the impact of dark energy-dark matter interaction with the \( \Lambda \)-CDM model. 
The cosmological parameter \( \Omega_m = 1 - x^2 - y^2 \) and its corresponding autonomous equation are: $\Omega_m' = -\alpha \left(\frac{1}{2} + \sqrt{\frac{3}{2}} x\right) \Omega_m + \frac{3}{2} (x^2 - y^2 + 1)(x^2 + y^2) - 3x^2$,
showing that \( \Omega_m \) depends on both \( x \) and \( y \). In the right panel of Fig.\ \ref{fig:FN2}, we present the \( \Omega_m - u \) phase space, with fixed points at \( x = 0.408 \) and \( y = 0.912 \), where this benchmark case (\( \alpha = 1, \lambda = 1 \)) shows stability.
Additionally, we used the autonomous equation for \( u \):
$u'= -u (u+2-\alpha-\sqrt{6}\alpha x) +\frac{3}{2} (1-x^2-y^2) +\frac{3}{2} (1+x^2-y^2) u$,
and analyzed the stability of the fixed point in this 2-D phase portrait.\\

 We identified two fixed points in the new phase space: one stable ($P_2$) and one saddle-type ($P_1$). All trajectories converge to the stable fixed point ($P_2$), where the growth rate is approximately 0.5 and \( \Omega_m = 0.017 \). In the same plot, we compare the behavior of the \( \Lambda \)-CDM model, where the growth rate is parametrized as \( u = \Omega_m^\gamma \) with \( \gamma = \frac{6}{11} \) (green line). At the present epoch (\( \Omega_m^0 \sim 0.28 \)), the growth rate is around 0.5, similar to that of the interacting quintessence model at the distant future, where \( \Omega_m \sim 0.0178 \). In the \( \Lambda \)-CDM model, the growth rate follows \( u = \Omega_m^\gamma \), tending to zero as \( \Omega_m \to 0 \). However, in the interacting model, the dark matter density in the source term ensures that \( u \to 0.5 \) at the stable fixed point ($P_2$), even as \( \Omega_m \to 0 \). This stable fixed point, in the dark-energy-dominated universe, mimics the growth rate behavior of the \( \Lambda \)-CDM model at late times. At the saddle fixed point (\( P_1 \)), all trajectories are repelled, representing zero growth rate at \( \Omega_m^0 = 0.0178 \). Thus, for \( \alpha = \lambda = 1 \), two fixed points are found: one stable with a positive growth rate and one saddle-type with a zero growth rate.

\section{Conclusion}
\label{con}
In this article, we investigated the impact of interactions between dark energy and dark matter on the formation of structure in the universe by analyzing the evolution of scalar perturbations over a flat spacetime background. 
We treat dark matter as non-relativistic (pressureless) dust with an equation of state, $\omega_{\rm m}=0$. Additionally, we consider the dynamics of dark energy to be driven by a quintessence scalar field $\phi$ with a potential given by 
$V(\phi) = V_0 e^{-\lambda\kappa\phi}$, where $V_0>0$ and $\lambda$ is a dimensionless parameter.
We incorporated interactions between the dark matter and dark energy components by employing a phenomenological model of interaction. This involved introducing a source term $\bar{Q}_0$ at the background level into the respective continuity equations of the two sectors with opposite signs, ensuring energy conservation within the combined dark matter and dark energy sector. Simultaneously, this facilitates the possibility of a continuous energy exchange between the two sectors at a rate determined by $\bar{Q}_0$.
We adopted a specific form of the source term,
$\bar{Q}_0 = \alpha \bar{\rho}_{\rm m}(H + \kappa \dot{\phi})$, 
to explore the evolutionary dynamics of the universe along with dark matter density
perturbations. 
We have neglected perturbations to the quintessence dark energy field $\phi$ on scales smaller than the horizon, which is a reasonable assumption during the era of growth
of matter perturbations.
The time derivative of the quintessence field ($\dot{\phi}$),
which appears in the adopted form of the source term $\bar{Q}_0$,
involves the quintessence-driven dark energy density $\bar{\rho}_\phi$.
This triggers the influence of dark energy density on the evolution of matter density perturbations within the considered model of interactions.
Consequently, it motivates investigating the potential of such interaction forms to address the cosmic coincidence problem.  
The coupling parameter $\alpha$ and the quintessence potential parameter $\lambda$ are pivotal in governing the dynamics of evolution explored within the framework of the interacting DE-DM scenario considered here. \\

Incorporation of DE-DM interactions results in modification of the Friedmann equations and the equation of motion of the scalar field at the background level, as well as the continuity and Euler equations in the matter sector at the perturbation level. Utilizing these equations along with Poisson's equation and assuming a negligible time variation of the gravitational potential leads to a single second-order differential equation for the DM 
density contrast $\delta_{\rm m}$.
We introduced a variable $u$ in the context of perturbation, defined as $u \equiv \frac{d(\ln \delta_{\rm m})}{d(\ln a)}$, which acts as a tracker of the growth of structure during the matter-dominated phase of the universe.
Additionally, we introduced two other dimensionless dynamical variables, $x$ and $y$, which respectively involve $\dot{\phi}$ and $V(\phi)$,   aiming to capture dynamical features associated with background evolution.
To analyze the evolution of perturbations using the dynamical analysis approach, we convert the evolution equations into a first-order autonomous system of equations
involving the three dynamical variables $x,y$ and $u$.
Additionally, the background cosmological quantities such as the  critical energy density parameters,
the grand EoS parameter, the deceleration parameter, and the coincidence parameter $r_{\rm mc}$ can be expressed in terms of these three dynamical variables.\\

We identified a total of 14 critical points within the autonomous system. We obtained constraints on the model parameters $\alpha$ and $\lambda$ for each of these critical points, ensuring their associated $(x, y, u)$ values are real, thereby confirming their existence. 
We extensively examined each fixed point, analyzing their stability criteria and determining whether they correspond to a matter, scaling, or dark energy (DE) dominance. We also investigated both accelerating and non-accelerating dynamics, along with the growth rate of matter perturbations, which may decay, remain constant, or grow over time. 
The stability criteria and other characteristics associated with the critical points depend on the model parameters $\alpha$ and $\lambda$. Thus, specifying the characteristics 
of the fixed points entails establishing pertinent ranges for the model parameters.
In obtaining the parameter ranges within which fixed points can exhibit
 stable, non-phantom type accelerating solutions, 
 we imposed the constraint $-1<\omega_{\rm tot}<-\frac13$ in addition to the viability condition $0<\Omega_{\rm m} < 1$.
  A scan of the $\alpha-\lambda$ parameter space reveals that both of the above constraints are simultaneously satisfied across a $\lambda$ range of [1-1.4] for any value of $\alpha$ between 0 and 1. Within this range of $\lambda$, as $\lambda$ approaches 1, the value of $\omega_{\rm tot}$ approaches $-1$ (the phantom barrier), mimicking the behaviour of $\Lambda$-CDM cosmology. 
This motivates the selection of benchmark values for the parameters ($\alpha = 0.01, 1$ and $\lambda = 1$) to illustrate the results of the dynamical analysis performed in the context of this work.\\

The value of the coupling parameter $\alpha$ has a significant impact on the characteristics of critical points. When the coupling is small ($\alpha = 0.01$), the critical point $G$ (in Table \ref{tab:T1}) behaves as an attractor. However, with moderate coupling ($\alpha = 1$), the same critical point shifts to being a repeller. At this moderate coupling, the critical point $E_+$ behaves as an attractor, a behavior opposite to its role as a repeller under small coupling ($\alpha = 0.01$).
 For the two chosen  values of $\alpha = 0.01 (1)$ with $\lambda$ fixed at 1, four (five) critical points have been identified,
and the remaining points either become complex or
 ail to meet the viability condition $0 < \Omega_{\rm m} < 1$.
Our investigation reveals the existence of no critical points at infinity.
Furthermore, we have depicted the trajectories in phase space around the fixed points corresponding to the chosen benchmark values of the model parameters.
We have identified  global attractors (referred to as point $G$ for $\alpha = 0.01$ and $E_+$ for $\alpha =1$) which satisfy all viability conditions and are capable of producing 
non-phantom-type accelerating solutions. The other fixed points are found to be 
 saddle points exhibiting the matter-dominated phase with either increasing or decreasing rates of growth of matter perturbations. \\
 
  A more profound and intriguing understanding of the interacting DE-DM scenario 
  can be achieved through the examination of the evolutionary profiles of various cosmological quantities.
We  found the profile of DE and DM energy density parameters ($\Omega_{\rm m}$
and $\Omega_{\rm \phi}$)
and hence also the coincidence parameter 
$r_{\text{mc}}\equiv \Omega_{\rm m}/\Omega_{\rm \phi}$ are significantly
depend on the coupling parameter $\alpha$. 
We observe a stronger dominance of DE over DM with higher $\alpha$ values
around the present epoch. 
A higher $\alpha$ indicates a more pronounced interaction between these DE and DM 
components, implying a greater rate of energy transfer from DM to DE. 
Consequently, in this model of interaction the value of $\alpha$ 
controls  the coincidence parameter 
$r_{\rm mc}\equiv \Omega_{\rm m}/\Omega_{\rm \phi}$
which is an indicator of the degree of alignment between dark matter and dark energy densities at present epoch.
Hence, the issue of cosmic coincidence can be addressed 
within the framework of the phenomenological model of interaction between
DE and DM sectors as examined in this study. In both weak and moderate coupling scenarios, a stable non-phantom dark energy era has been achieved within the current framework of the interacting scenario. \\

 Finally, we also examined the evolution of the growth rate of matter perturbations
 exploring the evolution of the perturbation variable $u$. 
For $\alpha = 0$ and $0.01$, we observed a significant fluctuation in the amplitude of the growth rate $u$, encompassing relatively large negative and positive values across the entire evolutionary phase. 
It increases and reaches its peak during the matter-dominated phase, 
subsequently declining gradually as the universe enters its acceleration phase.
 In scenarios with moderate interactions ($\alpha = 1$), the growth rate remains significantly suppressed but does not sharply drop; instead, it asymptotically approaches zero. Therefore, with moderate DE-DM interaction
 strength, the growth of matter perturbation persists not only until the onset of 
 the late-time epoch but may extend far into the future.\\

 To constrain the parameters $\alpha$ and $\lambda$ in an interacting dark energy (DE) and dark matter (DM) model, one may use Markov Chain Monte Carlo (MCMC) with the \texttt{emcee} Python package. The model is defined by the coupling term \(Q = \alpha \bar{\rho}_m (H + \kappa \dot{\phi})\) and the scalar field potential \(V(\phi) = V_0 e^{-\lambda \kappa \phi}\), where $\alpha$ and $\lambda$ control the strength of the DE-DM interaction and the steepness of the potential, respectively. Cosmological observables, such as the Hubble parameter \(H(z)\), are computed using a cosmology code like \texttt{emcee}. The likelihood function compares the model predictions with observational data (e.g., from Pantheon+, DESI, and Planck), while the prior function imposes physical constraints on $\alpha$ and $\lambda$. 
The results yield posterior constraints on $\alpha$ and $\lambda$, providing insights into the DE-DM interaction and identifying the best-fit values along with their uncertainties based on observational data.\\

However, evaluating these parameters from an observational standpoint necessitates a thorough and extensive analysis of observational datasets, which falls outside the focus and scope of this study. In our current work, we only focus primarily on the dynamical analysis approach to place constraints on the model parameters (\(\alpha\), \(\lambda\)). Both values of \(\alpha\) have been examined within this framework in fig. 3. Where, we presented a density plot showing the variation of the total EoS parameter, \(\omega_{\text{tot}}\), in the \((\alpha, \lambda)\) plane, subject to the constraint \(0 < \Omega_m < 1\). Within the range \(1 < \lambda < 2.3\), \(\omega_{\text{tot}}\) deviates from 1. Beyond this range, \(\omega_{\text{tot}}\) remains constant at 1, indicating a stiff matter phase. Notably, for \(1 < \lambda < 1.4\), \(\omega_{\text{tot}}\) lies between \(-1\) and \(-1/3\), signifying an accelerated expansion phase. These ranges of \(\lambda\) were determined by scanning \(\alpha\) values from 0 to 1. Given that the accelerating region is a primary focus of our study, we selected benchmark values \(\alpha = 0.01, 1\) and \(\lambda = 1\) for further analysis of phase space trajectories and the evolutionary dynamics within this interacting DE-DM system. No significant late-time acceleration or viable energy density constraints (\(0 < \Omega_m < 1\)) were found in the region of negative \(\alpha\) and \(\lambda\), leading to the exclusion of these regions from the plot in Fig.~3.
We intend to address this aspect in future work, where we will explore the implications of observational data on the coupling parameter \(\alpha\) and potential parameter $\lambda$.
  \\

In summary, the phenomenological model depicting the interaction between dark matter and the quintessence dark energy field $\phi$, as explored in the paper, effectively captures the evolutionary features of the universe across its various phases at both the background and perturbation levels. 
The DE-DM coupling parameter  facilitates the tuning of evolutionary features at both   
levels. 
Accordingly, within  appropriate parameter ranges, it becomes feasible to precisely model the evolutionary history of the universe, encompassing a matter-dominated phase with a natural transition into the dark energy-dominated non-phantom type accelerating phase. 
This  highlights the substantial impact of the interaction between dark energy and dark matter on the evolution of the universe's background spacetime and its matter perturbations. 
While not within the purview of the present study, exploring scenarios wherein the interactions between the Dark Energy  and Dark Matter  sectors stem from diverse perspectives holds promise for future investigations. This involves scrutinizing different types of couplings, such as curvature-matter and curvature-field-matter scenarios. Additionally, employing dynamical system analysis to explore the tensor mode of perturbations could offer deeper insights.

\paragraph{Acknowledgement}\
All authors would like to thank the referee for the valuable suggestions.


\begin{thebibliography}{101}

\bibitem{ref:Riess98}   
 A.~G.~Riess {\it et al.} [Supernova Search Team],
  Astron.\ J.\  {\bf 116}, 1009 (1998)
  doi:10.1086/300499
  [astro-ph/9805201].

\bibitem{ref:Perlmutter} 
S.~Perlmutter {\it et al.} [Supernova Cosmology Project Collaboration],
  Astrophys.\ J.\  {\bf 517}, 565 (1999)
  doi:10.1086/307221
  [astro-ph/9812133].



\bibitem{WMAP:2003elm}
D.~N.~Spergel \textit{et al.} [WMAP],
Astrophys. J. Suppl. \textbf{148} (2003), 175-194
doi:10.1086/377226
[arXiv:astro-ph/0302209 [astro-ph]]


\bibitem{Hinshaw:2008kr}
G.~Hinshaw \textit{et al.} [WMAP],
Astrophys. J. Suppl. \textbf{180} (2009), 225-245
doi:10.1088/0067-0049/180/2/225
[arXiv:0803.0732 [astro-ph]].


\bibitem{SDSS:2005xqv}
D.~J.~Eisenstein \textit{et al.} [SDSS],
Astrophys. J. \textbf{633} (2005), 560-574
doi:10.1086/466512
[arXiv:astro-ph/0501171 [astro-ph]]

\bibitem{mps1} S.~Cole \textit{et al.}   
 Mon. Not. Roy. Astron. Soc., 362:
505–534, 2005. doi:10.1111/j.1365-2966.2005.09318.x.


\bibitem{mps2} W.~J.~Percival  \textit{et al.}  
 Astrophys. J., 657:645–663, 2007. doi:
10.1086/510615.





\bibitem{Sofue:2000jx}
Y.~Sofue and V.~Rubin,
Ann. Rev. Astron. Astrophys. \textbf{39} (2001), 137-174
doi:10.1146/annurev.astro.39.1.137
[arXiv:astro-ph/0010594 [astro-ph]]


\bibitem{Bartelmann:1999yn}
M.~Bartelmann and P.~Schneider,
Phys. Rept. \textbf{340} (2001), 291-472
doi:10.1016/S0370-1573(00)00082-X
[arXiv:astro-ph/9912508 [astro-ph]]


\bibitem{Clowe:2006eq}
D.~Clowe, M.~Bradac, A.~H.~Gonzalez, M.~Markevitch, S.~W.~Randall, C.~Jones and D.~Zaritsky,
Astrophys. J. Lett. \textbf{648} (2006), L109-L113
doi:10.1086/508162
[arXiv:astro-ph/0608407 [astro-ph]]

\bibitem{Planck:2018vyg}
N.~Aghanim \textit{et al.} [Planck],
Astron. Astrophys. \textbf{641} (2020), A6
[erratum: Astron. Astrophys. \textbf{652} (2021), C4]
doi:10.1051/0004-6361/201833910
[arXiv:1807.06209 [astro-ph.CO]]










\bibitem{Zlatev:1998tr}
I.~Zlatev, L.~M.~Wang and P.~J.~Steinhardt,
Phys. Rev. Lett. \textbf{82} (1999), 896-899
doi:10.1103/PhysRevLett.82.896
[arXiv:astro-ph/9807002 [astro-ph]]






\bibitem{Martin:2012bt}
J.~Martin,
Comptes Rendus Physique \textbf{13} (2012), 566-665
doi:10.1016/j.crhy.2012.04.008
[arXiv:1205.3365 [astro-ph.CO]].


		
\bibitem{Peccei:1987mm}
R.~D.~Peccei, J.~Sola and C.~Wetterich,
Phys. Lett. B \textbf{195} (1987), 183-190
doi:10.1016/0370-2693(87)91191-9
		
\bibitem{Ford:1987de}
L.~H.~Ford,
Phys. Rev. D \textbf{35} (1987), 2339
doi:10.1103/PhysRevD.35.2339
		
\bibitem{Peebles:2002gy}
P.~J.~E.~Peebles and B.~Ratra,
Rev. Mod. Phys. \textbf{75} (2003), 559-606
doi:10.1103/RevModPhys.75.559
[arXiv:astro-ph/0207347 [astro-ph]].
		
\bibitem{Nishioka:1992sg}
T.~Nishioka and Y.~Fujii,
Phys. Rev. D \textbf{45} (1992), 2140-2143
doi:10.1103/PhysRevD.45.2140
		
\bibitem{Ferreira:1997au}
P.~G.~Ferreira and M.~Joyce,
Phys. Rev. Lett. \textbf{79} (1997), 4740-4743
doi:10.1103/PhysRevLett.79.4740
[arXiv:astro-ph/9707286 [astro-ph]].
		
\bibitem{Ferreira:1997hj}
P.~G.~Ferreira and M.~Joyce,
Phys. Rev. D \textbf{58} (1998), 023503
doi:10.1103/PhysRevD.58.023503
[arXiv:astro-ph/9711102 [astro-ph]]
		
\bibitem{Caldwell:1997ii}
R.~R.~Caldwell, R.~Dave and P.~J.~Steinhardt,
Phys. Rev. Lett. \textbf{80} (1998), 1582-1585
doi:10.1103/PhysRevLett.80.1582
[arXiv:astro-ph/9708069 [astro-ph]].
		
\bibitem{Carroll:1998zi}
S.~M.~Carroll,
Phys. Rev. Lett. \textbf{81} (1998), 3067-3070
doi:10.1103/PhysRevLett.81.3067
[arXiv:astro-ph/9806099 [astro-ph]]
		
\bibitem{Copeland:1997et}
E.~J.~Copeland, A.~R.~Liddle and D.~Wands,
Phys. Rev. D \textbf{57} (1998), 4686-4690
doi:10.1103/PhysRevD.57.4686
[arXiv:gr-qc/9711068 [gr-qc]].




\bibitem{Fang:2014qga}
W.~Fang, H.~Tu, Y.~Li, J.~Huang and C.~Shu,
Phys. Rev. D \textbf{89} (2014) no.12, 123514
doi:10.1103/PhysRevD.89.123514
[arXiv:1406.0128 [gr-qc]].
		
		
\bibitem{ArmendarizPicon:1999rj}
C.~Armendariz-Picon, T.~Damour and V.~F.~Mukhanov,
Phys. Lett. B \textbf{458} (1999), 209-218
doi:10.1016/S0370-2693(99)00603-6
[arXiv:hep-th/9904075 [hep-th]]
		
\bibitem{ArmendarizPicon:2000ah}
C.~Armendariz-Picon, V.~F.~Mukhanov and P.~J.~Steinhardt,
Phys. Rev. D \textbf{63} (2001), 103510
doi:10.1103/PhysRevD.63.103510
[arXiv:astro-ph/0006373 [astro-ph]]
		
\bibitem{ArmendarizPicon:2000dh}
C.~Armendariz-Picon, V.~F.~Mukhanov and P.~J.~Steinhardt,
Phys. Rev. Lett. \textbf{85} (2000), 4438-4441
doi:10.1103/PhysRevLett.85.4438
[arXiv:astro-ph/0004134 [astro-ph]]
		
\bibitem{ArmendarizPicon:2005nz}
C.~Armendariz-Picon and E.~A.~Lim,
JCAP \textbf{08} (2005), 007
doi:10.1088/1475-7516/2005/08/007
[arXiv:astro-ph/0505207 [astro-ph]]
		
\bibitem{Chiba:1999ka}
T.~Chiba, T.~Okabe and M.~Yamaguchi,
Phys. Rev. D \textbf{62} (2000), 023511
doi:10.1103/PhysRevD.62.023511
[arXiv:astro-ph/9912463 [astro-ph]]
		
\bibitem{ArkaniHamed:2003uy}
N.~Arkani-Hamed, H.~C.~Cheng, M.~A.~Luty and S.~Mukohyama,
JHEP \textbf{05} (2004), 074
doi:10.1088/1126-6708/2004/05/074
[arXiv:hep-th/0312099 [hep-th]]
		
\bibitem{Caldwell:1999ew}
R.~R.~Caldwell,
Phys. Lett. B \textbf{545} (2002), 23-29
doi:10.1016/S0370-2693(02)02589-3
[arXiv:astro-ph/9908168 [astro-ph]]




\bibitem{Bandyopadhyay:2017igc}
A.~Bandyopadhyay and A.~Chatterjee,
Mod. Phys. Lett. A \textbf{34} (2019) no.27, 1950219
doi:10.1142/S0217732319502195
[arXiv:1709.04334 [gr-qc]]

\bibitem{Bandyopadhyay:2018zlz}
A.~Bandyopadhyay and A.~Chatterjee,
Eur. Phys. J. Plus \textbf{134} (2019) no.4, 174
doi:10.1140/epjp/i2019-12587-0
[arXiv:1808.05259 [gr-qc]]

\bibitem{Bandyopadhyay:2019ukl}
A.~Bandyopadhyay and A.~Chatterjee,
Eur. Phys. J. Plus \textbf{135} (2020) no.2, 181
doi:10.1140/epjp/s13360-020-00161-w
[arXiv:1902.04315 [gr-qc]]

\bibitem{Bandyopadhyay:2019vdd}
A.~Bandyopadhyay and A.~Chatterjee,
Res. Astron. Astrophys. \textbf{21} (2021) no.1, 002
doi:10.1088/1674-4527/21/1/2
[arXiv:1910.10423 [gr-qc]]

\bibitem{Chatterjee:2022uyw}
A.~Chatterjee, B.~Jana and A.~Bandyopadhyay,
Eur. Phys. J. Plus \textbf{137} (2022) no.11, 1271
doi:10.1140/epjp/s13360-022-03476-y
[arXiv:2207.00888 [gr-qc]]




\bibitem{fr1}
S.~Capozziello,
Int. J. Mod. Phys. D \textbf{11} (2002), 483-492
doi:10.1142/S0218271802002025
[arXiv:gr-qc/0201033 [gr-qc]]

\bibitem{fr2}
S.~Capozziello, V.~F.~Cardone, S.~Carloni and A.~Troisi,
Int. J. Mod. Phys. D \textbf{12} (2003), 1969-1982
doi:10.1142/S0218271803004407
[arXiv:astro-ph/0307018 [astro-ph]]


\bibitem{fr3}
S.~Nojiri and S.~D.~Odintsov,
Phys. Rev. D \textbf{68} (2003), 123512
doi:10.1103/PhysRevD.68.123512
[arXiv:hep-th/0307288 [hep-th]]

\bibitem{fr4}
S.~Nojiri and S.~D.~Odintsov,
Phys. Rept. \textbf{505} (2011), 59-144
doi:10.1016/j.physrep.2011.04.001
[arXiv:1011.0544 [gr-qc]]

\bibitem{fr5}
S.~Nojiri, S.~D.~Odintsov and V.~K.~Oikonomou,
Phys. Rept. \textbf{692} (2017), 1-104
doi:10.1016/j.physrep.2017.06.001
[arXiv:1705.11098 [gr-qc]]
  



\bibitem{fr6}
B.~Jana, A.~Chatterjee, K.~Ravi and A.~Bandyopadhyay,
Class. Quant. Grav. \textbf{40} (2023) no.19, 195023
doi:10.1088/1361-6382/acf554
[arXiv:2303.06961 [gr-qc]].


\bibitem{fr7}
K.~Ravi, A.~Chatterjee, B.~Jana and A.~Bandyopadhyay,
MNRAS, \textbf{527}, 3, (January 2024), Pages 7626–7651
DOI: 10.1093/mnras/stad3705
[arXiv:2306.12585 [astro-ph.CO]].

\bibitem{fr8}
A.~Chatterjee, R.~Roy, S.~Dey and A.~Bandyopadhyay,
Eur. Phys. J. C \textbf{84} (2024) no.3, 236
doi:10.1140/epjc/s10052-024-12611-1
[arXiv:2310.05578 [gr-qc]]

\bibitem{fr9}
A.~Chatterjee,
Class. Quant. Grav. \textbf{41} (2024) no.9, 095007
doi:10.1088/1361-6382/ad2f11
[arXiv:2312.07173 [gr-qc]]


\bibitem{mgst1}  A.~S.~Bhatia and S.~Sur,   Int.~J.~Mod.~Phys.~D 26 (2017) 1750149, [arXiv:1702.01267 [gr-qc]].
 
\bibitem{mgst2}  Y.~Fujii and K.~Maeda, The Scalar-Tensor Theory of Gravitation, Cambridge Monographs on
Mathematical Physics, Cambridge University Press, United Kingdom (2003).
 
\bibitem{mgst3}  N.~Bartolo and M.~Pietroni,   Phys.~Rev.~D 61 (2000)
023518, [hep-ph/9908521].
 
 
\bibitem{mgst4} S.~Tsujikawa, K.~Uddin, S.~Mizuno, R.~Tavakol and J.~I.~Yokoyama, Phys.~Rev.~D 77 (2008) 103009. 

\bibitem{Chimento:2003iea}
L.~P.~Chimento, A.~S.~Jakubi, D.~Pavon and W.~Zimdahl,
Phys. Rev. D \textbf{67} (2003), 083513
doi:10.1103/PhysRevD.67.083513
[arXiv:astro-ph/0303145 [astro-ph]]





\bibitem{ht1} N.~Aghanim \textit{et al.} 
Astron.~Astrophys.,~ 641:A6,
2020. doi:10.1051/0004-6361/201833910. [Erratum: Astron.Astrophys. 652, C4 (2021)].

\bibitem{ht2}  Adam~G.~Riess \textit{et al.}   Astrophys.~J.~Lett., 934(1):L7, 2022.
doi:10.3847/2041-8213/ac5c5b.





\bibitem{An:2017crg}
R.~An, C.~Feng and B.~Wang,
JCAP \textbf{02} (2018), 038
doi:10.1088/1475-7516/2018/02/038
[arXiv:1711.06799 [astro-ph.CO]].
\bibitem{Pourtsidou:2016ico}
A.~Pourtsidou and T.~Tram,
Phys. Rev. D \textbf{94} (2016) no.4, 043518
doi:10.1103/PhysRevD.94.043518
[arXiv:1604.04222 [astro-ph.CO]].




\bibitem{Kumar:2017dnp}
S.~Kumar and R.~C.~Nunes,
Phys. Rev. D \textbf{96} (2017) no.10, 103511
doi:10.1103/PhysRevD.96.103511
[arXiv:1702.02143 [astro-ph.CO]].



\bibitem{kids}  H.~Hildebrandt   \textit{et al.},
 Mon. Not. Roy. Astron. Soc., 465 (2017) 1454,
[arXiv:1606.05338].

\bibitem{plxi}  N.~Aghanim   \textit{et al.}, 
Astron.Astrophys., 594 (2016) A11, [arXiv:1507.02704].


\bibitem{Yang:2018uae}
W.~Yang, A.~Mukherjee, E.~Di Valentino and S.~Pan,
Phys. Rev. D \textbf{98} (2018) no.12, 123527
doi:10.1103/PhysRevD.98.123527
[arXiv:1809.06883 [astro-ph.CO]].



\bibitem{DiValentino:2021izs}
E.~Di Valentino, O.~Mena, S.~Pan, L.~Visinelli, W.~Yang, A.~Melchiorri, D.~F.~Mota, A.~G.~Riess and J.~Silk,
Class. Quant. Grav. \textbf{38} (2021) no.15, 153001
doi:10.1088/1361-6382/ac086d
[arXiv:2103.01183 [astro-ph.CO]]


\bibitem{qft1} A.~A.~Costa, L.~C.~Olivari and E.~Abdalla,   Phys. Rev. \textbf{D92}  (2015) 103501 [arXiv:1411.3660]

\bibitem{qft2} G.~R.~Farrar and P.~J.~E.~Peebles,   Astrophys. J. \textbf{604}
(2004) 1 [astro-ph/0307316]





\bibitem{Wetterich:1994bg}
C.~Wetterich,
Astron. Astrophys. \textbf{301} (1995), 321-328
[arXiv:hep-th/9408025 [hep-th]].


\bibitem{Amendola:1999er}
L.~Amendola,
Phys. Rev. D \textbf{62} (2000), 043511
doi:10.1103/PhysRevD.62.043511
[arXiv:astro-ph/9908023 [astro-ph]].



\bibitem{Bernardi:2016xmb}
F.~F.~Bernardi and R.~G.~Landim,
Eur. Phys. J. C \textbf{77} (2017) no.5, 290
doi:10.1140/epjc/s10052-017-4858-x
[arXiv:1607.03506 [gr-qc]].

\bibitem{Duniya:2013eta}
D.~Duniya, D.~Bertacca and R.~Maartens,
JCAP \textbf{10} (2013), 015
doi:10.1088/1475-7516/2013/10/015
[arXiv:1305.4509 [astro-ph.CO]].

\bibitem{Marcondes:2016reb}
R.~J.~F.~Marcondes, R.~C.~G.~Landim, A.~A.~Costa, B.~Wang and E.~Abdalla,
JCAP \textbf{12} (2016), 009
doi:10.1088/1475-7516/2016/12/009
[arXiv:1605.05264 [astro-ph.CO]].



\bibitem{Dutta:2017wfd}
J.~Dutta, W.~Khyllep and N.~Tamanini,
JCAP \textbf{01} (2018), 038
doi:10.1088/1475-7516/2018/01/038
[arXiv:1707.09246 [gr-qc]].

\bibitem{Khyllep:2021wjd}
W.~Khyllep, J.~Dutta, S.~Basilakos and E.~N.~Saridakis,
Phys. Rev. D \textbf{105} (2022) no.4, 043511
doi:10.1103/PhysRevD.105.043511
[arXiv:2111.01268 [gr-qc]]





\bibitem{Chatterjee:2021ijw}
A.~Chatterjee, S.~Hussain and K.~Bhattacharya,
Phys. Rev. D \textbf{104} (2021) no.10, 2021
doi:10.1103/PhysRevD.104.103505
[arXiv:2105.00361 [gr-qc]]


\bibitem{Chatterjee:2021hhj}
A.~Chatterjee, A.~Bandyopadhyay and B.~Jana,
Eur. Phys. J. Plus \textbf{137} (2022) no.4, 518
doi:10.1140/epjp/s13360-022-02747-y
[arXiv:2108.12186 [gr-qc]]


\bibitem{Hussain:2022osn}
S.~Hussain, A.~Chatterjee and K.~Bhattacharya,
Universe \textbf{9} (2023) no.2, 65
doi:10.3390/universe9020065
[arXiv:2203.10607 [gr-qc]]


\bibitem{Bhattacharya:2022wzu}
K.~Bhattacharya, A.~Chatterjee and S.~Hussain,
Eur. Phys. J. C \textbf{83} (2023) no.6, 488
doi:10.1140/epjc/s10052-023-11666-w
[arXiv:2206.12398 [gr-qc]]

\bibitem{Hussain:2023kwk}
S.~Hussain, A.~Chatterjee and K.~Bhattacharya,
Phys. Rev. D \textbf{108} (2023) no.10, 103502
doi:10.1103/PhysRevD.108.103502
[arXiv:2305.19062 [gr-qc]]

 
\bibitem{Chatterjee:2023uga}
A.~Chatterjee, S.~Hussain and K.~Bhattacharya,
Springer Proc. Phys. \textbf{304} (2024), 303-306
doi:10.1007/978-981-97-0289-3\_66
[arXiv:2307.04633 [gr-qc]]



\bibitem{Cabral:2009hoy}
G.~Caldera-Cabral, R.~Maartens and B.~M.~Schaefer,
JCAP \textbf{07} (2009), 027
doi:10.1088/1475-7516/2009/07/027
[arXiv:0905.0492 [astro-ph.CO]]

\bibitem{Tsujikawa:2012hv}
S.~Tsujikawa, A.~De Felice and J.~Alcaniz,
JCAP \textbf{01} (2013), 030
doi:10.1088/1475-7516/2013/01/030
[arXiv:1210.4239 [astro-ph.CO]].


\bibitem{Chatterjee:2024rbh}
A.~Chatterjee, A.~Panda and A.~Bandyopadhyay,
Class. Quant. Grav. \textbf{42} (2025), 025019
doi:10.1088/1361-6382/ad9f18
[arXiv:2407.14791 [gr-qc]] 









































































\end{thebibliography}
\end{document}